\documentclass[prd,twocolumn,superscriptaddress,amsmath,amssymb,showpacs]{revtex4}

\usepackage{color}
\usepackage{times}
\usepackage{fancyhdr}
\usepackage{float}
\usepackage{alltt}
\usepackage{listings}
\usepackage{enumerate}
\usepackage[T1]{fontenc}
\usepackage{ae}
\usepackage{rotating}
\usepackage{subfigure}
\usepackage{amsmath}
\usepackage{amssymb}
\usepackage{graphicx}
\usepackage{dcolumn}
\usepackage{bm}
\usepackage{aas_macros_pk}

\def\be{\begin{equation}}
\def\ee{\end{equation}}
\def\beg{\begin{align}}
\def\eeg{\end{align}}
\def\bi{\begin{itemize}}
\def\ei{\end{itemize}}
\def\ben{\begin{enumerate}[1.]}
\def\een{\end{enumerate}}
\def\i{\item}

\newcommand{\abs}[1]{\left| {#1} \right|}

\newcommand{\bo}{\raise-1mm\hbox{\Large$\Box$}}
\newcommand{\sci}[2]{#1 \times 10^{#2}}
\newcommand{\rthz}{\mathrm{Hz}^{-\frac{1}{2}}}
\newcommand{\hrss}{h_{\mathrm{rss}}}
\newcommand{\hrssf}{h_{\mathrm{rss}}^{50\%}}
\newcommand{\hrssn}{h_{\mathrm{rss}}^{90\%}}

\newcommand{\egwn}{E_{\mathrm{GW}}^{90\%}}
\newcommand{\egw}{E_{\mathrm{GW}}}
\newcommand{\eem}{E_{\mathrm{EM}}}

\newcommand{\cc}[1]{{#1}^{\ast}}

\newcommand\ligodoc{P0900001}

\begin{document}

\title{Stacking Gravitational Wave Signals from Soft Gamma Repeater Bursts\\
}

\newcommand*{\CH}{California Institute of Technology, Pasadena, CA  91125, USA}
\affiliation{\CH}

\newcommand*{\CO}{Columbia University, New York, NY  10027, USA}
\affiliation{\CO}

\newcommand*{\PU}{
Institute for Gravitation and the Cosmos, Center for Gravitational Wave
Physics, and Department of Physics,
The Pennsylvania State University, University Park, PA  16802, USA}
\affiliation{\PU}

\author{P.~Kalmus}   \email[]{peter.kalmus@ligo.org} \affiliation{\CH} \affiliation{\CO} 

\author{K.~C.~Cannon}  \affiliation{\CH}

\author{S.~M\'{a}rka}    \affiliation{\CO} 

\author{B.~J.~Owen}    \affiliation{\PU}

\date{\today}

\begin{abstract}

Soft gamma repeaters (SGRs) have unique properties that make them
intriguing targets for gravitational wave (GW) searches.  
They are nearby, their burst emission  mechanism may involve neutron star
crust fractures and excitation of quasi-normal
modes, and they burst repeatedly and sometimes spectacularly.  A recent LIGO search for transient GW from  these 
sources placed upper limits on a set of almost 200 individual  SGR bursts.  These limits were within the theoretically 
predicted  range of some models.  We present a new search strategy which builds upon the method used there by
``stacking'' potential GW  signals from multiple SGR bursts.  We assume that variation in the  time difference between 
burst electromagnetic emission and burst GW emission is small relative to the  GW signal duration, and we 
time-align GW excess power time-frequency tilings containing individual burst triggers to their corresponding 
electromagnetic emissions.    Using Monte Carlo simulations, we confirm that gains in GW energy sensitivity of 
$N^{1/2}$ are  possible, where $N$ is the number of stacked SGR bursts.   Estimated sensitivities for a mock search 
for gravitational waves from the 2006 March 29 storm from SGR 1900+14 are also presented, for two GW emission models, ``fluence-weighted"  and  ``flat" (unweighted).

\end{abstract}


\maketitle

\section{Introduction}

Soft gamma repeaters (SGRs) are promising potential sources of
gravitational waves (GWs). They sporadically emit brief ($\approx0.1$\,s) intense
bursts of soft gamma-rays with peak luminosities commonly up to $10^{42}$
erg/s.  Three of the five known galactic SGRs have produced rare ``giant
flare'' events with initial bright, short ($\approx0.2$\,s) pulses followed
by tails lasting minutes, with peak luminosities between $10^{44}$ and
$10^{47}$\,erg/s.  According to the ``magnetar'' model SGRs are
neutron stars with extreme magnetic fields
$\sim\nolinebreak10^{15}$\,G\,\cite{duncan92}.  Bursts  may result from the
interaction of the star's magnetic field with its solid crust, leading to
crustal deformations and occasional catastrophic
cracking\,\cite{thompson95, schwartz05, horowitz09} with subsequent excitation of
nonradial star modes\,\cite{andersson97, pacheco98, ioka01} and the
emission of GWs\,\cite{owen05, horvath05, pacheco98, ioka01}.   For
reviews, see \,\cite{mereghetti08, woods04a}.

The LIGO Scientific Collaboration recently completed a search for transient gravitational waves associated with almost 
200 individual electromagnetic SGR triggers\,\cite{s5y1sgr}.  That search did not detect GW, but it explicitly targeted neutron star $f
$-modes, placed the most stringent upper limits on transient gravitational wave amplitudes at the time it was published, and set isotropic 
emission energy upper limits that fell within the theoretically predicted range of some SGR models\,\cite{ioka01}.

In this paper we extend that work and describe a new electromagnetically triggered search method for gravitational waves from multiple SGR bursts.  Triggered gravitational wave searches assume that gravitational waves associated with an electromagnetic astrophysical event would reach earth at approximately the same time as light from the event.  In addition, source sky location is also known.  

Knowledge of time and sky position can be a great advantage to
gravitational wave searches~\cite{Abbott:2008mr}.
It allows us to calculate the detector
response functions, allowing us to estimate more relevant upper limits on GW emission  using
simulated signals from the source.  Also, upper
limits are typically lower than for untriggered all-sky
searches, largely because they are more robust to loud glitches.
Finally, searches can target known astrophysical events.  If the distance to the source is known, 
results can be given in terms of isotropic gravitational wave energy
emitted from the source instead of strain amplitude at the detector.
This ties the search to the astrophysical source instead of the
detector on Earth.  All of these advantages apply to searches for
gravitational waves from SGR bursts.

The new analysis method described here, ``Stack-a-flare,'' builds upon the analysis pipeline (``Flare pipeline'') used in \,\cite{s5y1sgr} and described in\,
\cite{kalmus07, kalmus08}  by attempting to ``stack" potential gravitational wave signals from multiple SGR bursts.     To stack $N$ bursts, we first generate $N$ excess power time-frequency tilings.  These are  2-dimensional matrices in time and frequency generated from the two detectors' data streams.  Each tiling element gives an excess power estimate in the GW detector data stream in a small period of time $\delta t$ and a small range of frequency $\delta f$.  The time range of each tiling is chosen to be centered on the time of one of the target EM bursts in the storm.  We then align these $N$ tilings along the time dimension so that times of the target EM bursts coincide, and perform a weighted addition according to a particular GW emission model.

\begin{figure*}[!t]
\begin{center}
\includegraphics[angle=0,width=180mm, clip=false]{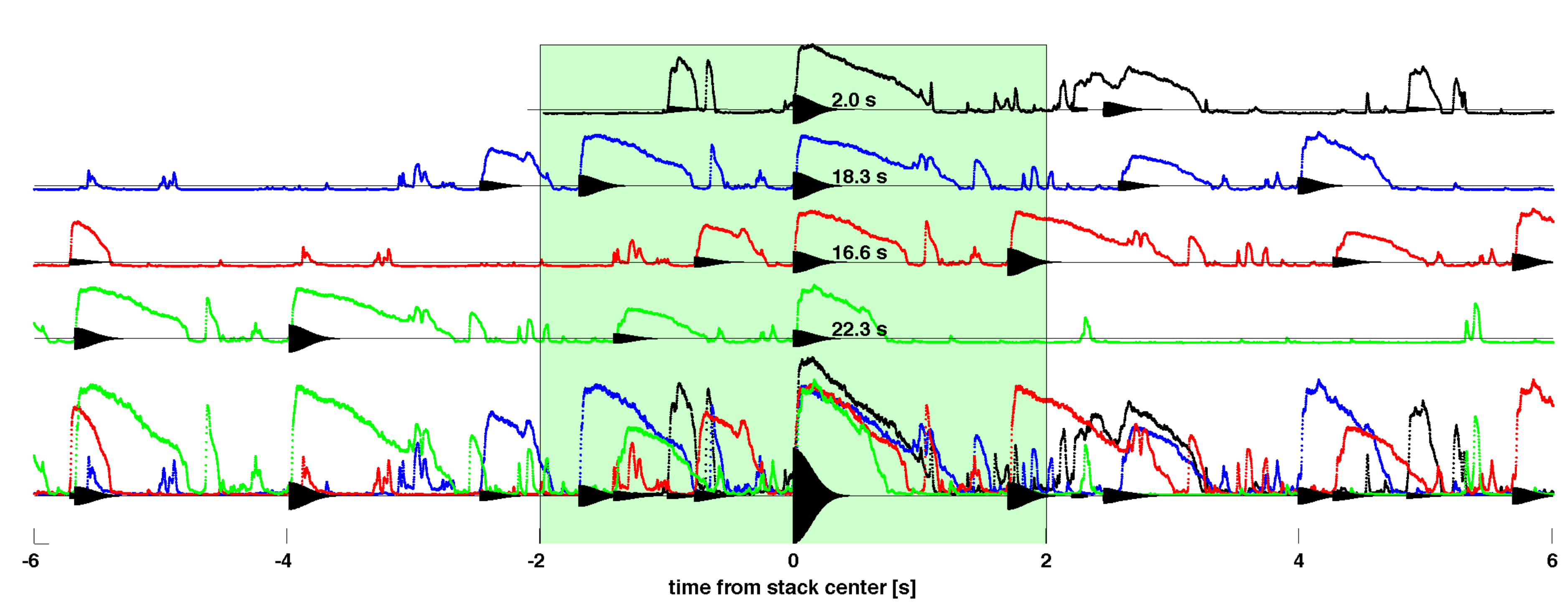}
\caption[] {Individual EM bursts inform the stacking of GW data.  This figure suggests the stacking procedure and explicitly shows search timescales.  The top four plots are 
EM light curve time series around individual bursts beginning at 2.0 s, 16.6 s, 18.3 s, and 22.3 s in Fig. \,\ref{fig:lc_rises}.  Simulated GW ringdowns in the fluence-weighted model are superposed.    The bottom plot shows the EM time series simultaneously, and the sum of the hypothetical GW signals.
The on-source region of $\pm2$\,s is shaded.    In a search, GW data corresponding to the EM time series are transformed to time-frequency power tilings before being added together and therefore there is no dependence on phase-coherence of GW signals in the analysis; this transformation is not illustrated.  } \label{fig:lc_stack}
\end{center}
\end{figure*}

We hope to improve the search sensitivity by combining potential gravitational wave signals from separate bursts in an attempt to 
increase the signal-to-noise ratio, increasing the probability of detection and placing more stringent constraints on 
theoretical models via upper limits.   We expect that this method would be suitable for performing 
searches using data from interferometric detectors such as LIGO's, for gravitational waves associated with SGR storm events such as the 2006 March 29 SGR 1900+14 storm.  Figure\,\ref{fig:lc_raw} shows the storm 
light curve obtained by the BAT detector aboard the Swift satellite\,\cite{BAT}, and Figure\,\ref{fig:lc_stack} illustrates the stacking procedure with the four most energetic bursts from the 
storm, showing the main search timescales.  

This paper is organized as follows.  In Section\,\ref{section:stackStrategy} we discuss the general strategy of  multiple 
SGR burst searches.   In Section\,\ref{section:stackMethod} we describe two versions of the analysis method (``Stack-a-flare''), both of which are built upon the Flare pipeline.   One version coherently stacks GW  time series 
associated with electromagnetic bursts in the storm, while the other stacks incoherently.  We  characterize the two 
versions using simulations in gaussian noise, demonstrating the strengths and weaknesses of each and showing that 
relatively weak signals which could not be detected in the individual burst search can easily be detected with the new method.  Gains 
in GW energy sensitivity of $N^{1/2}$ are feasible with the incoherent version even in the presence of realistic relative 
timing uncertainties between SGR bursts, where $N$ is the number of stacked SGR bursts.  Finally, in Section\,
\ref{section:stacksim} we examine the SGR 1900+14 storm light curve of Figure\,\ref{fig:lc_raw} in detail and present 
estimated search sensitivities for a simulated search for gravitational waves from the storm, for two GW emission models: 
``fluence-weighted"  and ``flat" (unweighted).

\section{Strategy} \label{section:stackStrategy}

Before discussing Stack-a-flare method implementation details, in this section we discuss the strategy of a generalized multiple SGR GW search.  Due to similarities between the 
individual burst and multiple burst searches in terms of astrophysics, goals, and implementation, the search choices 
will be similar to those followed in\,\cite{s5y1sgr}.

\subsection{Parameter space of the target GW signals}

As in the individual SGR search, multiple SGR searches could target
neutron star fundamental mode ringdowns (RDs) predicted in\,\cite{andersson97, pacheco98, ioka01, andersson02} as well as unknown
short-duration GW signals.
The former are interesting targets because fundamental modes are the modes most
efficiently damped by GW emission.
The latter are interesting targets because the flare mechanism may involve
some other violent motion inside the star, for example matter carried along
with magnetic field lines as they rearrange themselves inside the crust.    We assume that given a neutron star, $f$-mode frequencies and damping 
timescales would be similar from event to event, and that unknown signals would at least have similar central frequencies and durations from event to event. 

We thus focus on two distinct regions in the target signal time-frequency parameter space.  The first region targets  $\sim$(100-400)\,ms 
duration signals in the (1--3)\,kHz band, which includes $f$-mode RD signals  predicted in\,\cite{benhar04} for ten realistic neutron star 
equations of state.  We choose a search band of (1--3)\,kHz for RD 
searches, with a 250\,ms time window which was found to give optimal search sensitivity\,\citep{kalmus08}.  The second region targets $\sim$(5--200)\,ms duration signals in the (100--1000)\,Hz band. The target durations are set by prompt SGR burst timescales (5\,ms to 200\,ms) and the target frequencies are set by the detector's sensitive region.  We search 
in two bands: (100--200)\,Hz (probing the region in which the detectors are most sensitive) and (100--1000)\,Hz (for full spectral coverage below 
the RD search band) using a 125\,ms time window.  In all, we could search in three frequency bands. 

We could explore these areas in the parameter space using the same twelve simulated waveform types used for setting upper
limits in the individual SGR burst search, described in\,\cite{s5y1sgr,
kalmus08}:  linearly and circularly polarized RDs with $\tau=200$\,ms and
frequencies in the range 1--3\,kHz; and band- and time-limited white noise
bursts (WNBs) with durations of 11\,ms and 100\,ms and frequency bands
matched to the two low frequency search bands.  Polarization angle will be
chosen randomly for every compound injection in initial searches.

It seems plausible to assume that, for a given neutron star, $f$-mode
frequencies and damping timescales would be similar from event to event.
For pure $f$-modes these quantities depend only on the global structure of
the star, essentially the mean density~\cite{andersson97}, and the magnetic
corrections depend on integrals of the field over the entire stellar
interior (e.g.~\cite{Glampedakis:2007ga}).
However, the major motivation for the low frequency unknown-waveform portion of
the parameter space is stochastic gravitational wave emission arising from violent
events in the neutron star crust.  Therefore, we will not assume similar
waveforms from event to event in the unknown-waveform portion, although we will
assume similar central frequencies and durations.

\begin{figure*}[!t]
\includegraphics[angle=0,width=180mm,clip=false]{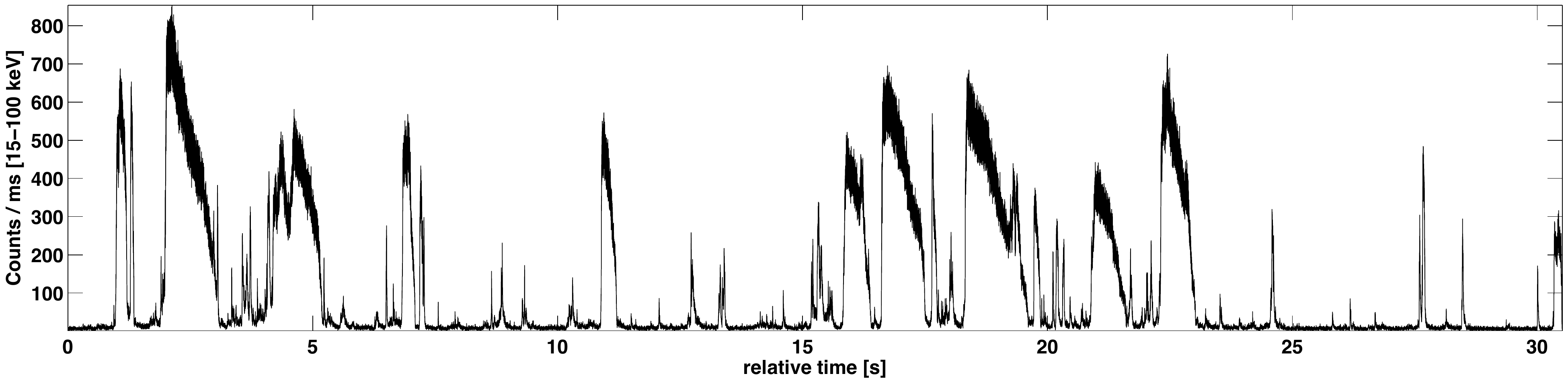}
\caption[BAT light curve of the SGR 1900+14 storm event] { BAT light curve of the SGR 1900+14 storm event, 1\,ms 
bins, 15-100\,keV.  The light curve shows the BAT event data, from sequence 00203127000, from approximately 20\,s 
after the start of the sequence to its end.  The data are publicly available\,\cite{swiftData}.   Several of the largest bursts,  durations greater than 500\,ms, are considered 
intermediate flares.  The x-axis times  are relative to 2006-03-29 02:53:09.9$\pm$0.5\,s UT at the satellite.    Times used in the analysis itself have been corrected for 
time-of-flight delays to the geocenter, which change by 0.35\,ms over the $\sim$30.5\,s course of the light curve.}
\label{fig:lc_raw}
\end{figure*}

\subsection{On-source region}

In a search, we divide the GW data into an on-source time region, in which GWs associated with a given burst 
could be expected, and a background time region.   On-source and background segments are analyzed identically 
resulting in lists of analysis events.

For the individual SGR search, trigger timing precision 
on the order of a second was handled with 4\,s on-source regions. For a 
multiple SGR search, significantly higher precision in \emph{relative} trigger times between burst events in the stack 
will be required.  However, a common bias in trigger times shared by all bursts in the stacking set (\emph{absolute} 
timing) can be handled with an adequately large (e.g. 4\,s) on-source region.

In general, imprecision in trigger times comes principally from two
sources: satellite to geocenter light crossing delay and arbitrariness of
the satellite trigger point in the light curve.  It is possible to decrease
both uncertainties through careful analysis of satellite data.  If
necessary, light crossing times at satellites can be propagated to the
geocenter (and subsequently to any given interferometer) using the
appropriate ephemeris.  If satellite data is public, light curve analysis
can yield a specific timing point, for example the start of the steep burst
rise.  Increased absolute timing precision could allow us to use smaller
on-source regions with durations set by theoretical predictions of time
delay between electromagnetic and gravitational wave emission from SGR
bursts.  However, this could exclude astrophysical models with larger timing delays, for
instance if the initial excitation transfers energy to the $f$-mode through
some slow and inefficient mode-mode coupling due to selection rules.
And it turns out there is little benefit to be gained:  We have performed
Monte Carlo simulations comparing $\pm2$\,s and $\pm1$\,s on-source
regions.
Reducing the on-source region from 4\,s to 2\,s resulted in only a 2\%
reduction in amplitude upper limits, on average over 24 trials with various
waveform types. 

We make one new assumption about the nature of individual bursts of
gravitational waves from SGRs: that the time delay between each gamma-ray
burst and the associated GW burst does not vary by more than about 100~ms,
less than duration of the shortest coherently modeled signal and comparable
to the star's Alfv\'en crossing time.
A variation this large would severely degrade detection efficiency in the case of 11\,ms duration WNBs, but as we shall see in Section\,\ref{section:sensitivityVsTiming}, detection of neutron star RDs (with $\tau \sim200$\,s) remains robust.

\subsection{Background region}

As with the individual burst search, the background region serves three purposes\,\cite{kalmus07, kalmus08}:
 \ben
 \i it is used to estimate statistics of the power tiling as a function of  frequency for use in the Flare pipeline; 
 \i it provides false alarm rate (FAR) estimates from which the significance of the loudest on-source analysis event can be determined;
  \i it is representative noise into which simulated waveforms can be injected for estimating upper limits.
 \een
In the course of validating the individual SGR search we showed that 1000\,s of data on either side of an on-source 
region produce sufficient estimates of the power tiling statistics\,\cite{kalmus08}.   This requirement and the estimation 
procedure are unchanged in the multiple SGR search, so $\pm1000$\,s of data will again suffice for this purpose.   
The background region required for injecting simulations to estimate upper limits may depend on the system being 
modeled and the desired statistical precision;  for the mock SGR 1900+14 storm search we describe below, $
\pm1000$\,s of background is sufficient.  The background region required for FAR estimates depends on the 
desired precision of FAR estimates.  Estimating the FAR of a very large on-source analysis event  requires a larger 
background than estimating the FAR of a small on-source analysis event, for a given level of precision.

\subsection{GW emission models}
\label{section:groupingAndWeighting}

In a multiple burst search we must choose which bursts to include in the
set and how to weight them.  Since GW energy  $\egw$ is unknown we are left to attempt to choose weightings via other observables which we hope will correlate to it.
As with the individual burst search, we assume
that the SGR burst sample is comprised of bursts occurring within some
specified time range defined by the observatory's science run schedule, and
not necessarily from the same SGR source.  We will refer to a set of SGR
bursts to be included in the multiple burst search, along with a weighting
strategy, as a GW emission model.

We could use Occam's razor to select  GW emission models such as the following:
 \begin{enumerate}[s1.]
 \i flat model --- use every detected and confirmed burst from a given SGR source within the time range, with equal weighting.  In practice we need to choose a cutoff on which bursts to include;
 \i inclusive model --- use every detected and confirmed burst within the time range, from any SGR source, with equal weighting;
 \i fluence-weighted model --- use every detected and confirmed burst from a given SGR source within the time range, weighted proportional to 
fluence;
 \i use a subset of component bursts from a multi-episodic burst event such as the SGR 1900+14 storm, with some 
weighting scheme.
 \end{enumerate}
 
We note that the inclusive model s2 could benefit from a search method that was insensitive to variations between 
SGR sources.  For example, we would expect two different sources to have $f$-modes at different frequencies, 
which may not brighten corresponding pixels in a time-frequency tiling.  We do
not attempt to solve this problem here.

We can also imagine stacking GW emission models based on specific theoretical
models.
A theory may predict that there is no correlation between
$\eem$ and $\egw$.  Such a prediction corresponds to the flat model and
is tested below.
Or a theory may predict that there is one emission mechanism with a
constant mechanical efficiency, and thus $\eem$ is always proportional to
$\egw$.
This is treated in the fluence-weighted model and tested below.
A theory may predict that $\egw$ is some more complicated function of
$\eem$, the simplest example being a step function---for instance, a theory
that small flares originate from magnetic reconnections far out in the
magnetosphere and only large flares from reconnections in or near the star
itself~\cite{lyutikov02}.
However the Swift/BAT spectra of the SGR~1900+14 storm indicate that the
neutron star surface  participates in all flares~\cite{israel08}, and
therefore we do not address such a model below (although it might be
interesting at a later date).
A theory may also predict that the time delay between electromagnetic and
gravitational emission varies from burst to burst.
Some variation, up to the target signal durations of tens or hundreds of
milliseconds, is accounted for in the searches described below.
Larger variations could potentially be treated by sweeping over some range
of time delays for each burst, a more complicated search which we do not
yet address.

We will neglect for the time being complicating factors such as: multiple
injections of energy into a single burst, with possible correlation in
gravitational wave emission energy;  qualitatively different gravitational
wave emission in the case of intermediate flares and common bursts; and
beaming issues.
We also neglect other possibilities such as correlating $\egw$ with flare
peak luminosities or rise times.

\section{Analysis method} \label{section:stackMethod}

Both versions of the Stack-a-flare pipeline, ``T-Stack'' and ``P-Stack,'' consist of an extension to the Flare 
pipeline\,\cite{kalmus08, kalmus07}, a simple but effective search pipeline based on the excess power detection 
statistic of~\cite{anderson01}.

\subsection{Flare pipeline}

The Flare pipeline can process streams of data from either one or two GW detectors.  First data is conditioned, then excess power tilings are created, and finally analysis events are constructed via clustering elements in the tilings.

Data conditioning consists of zero-phase digital filtering in the
time domain~\cite{hamming98}, first with a bandpass filter and then
with a composite notch filter. The raw calibrated LIGO power
spectrum is colored, and is characterized by a sensitive region
between $\sim$60~Hz to $\sim$2~kHz which includes a forest of narrow
lines, with increasingly loud noise on either side of the sensitive
band.  Search sensitivity is increased
by removing these insensitive regions from the data, which would
otherwise dominate weak signals and destroy bandwidth after
transformation to frequency domain.  We remove narrow lines associated with the power line
harmonics at multiples of 60~Hz, the violin modes of the mirror
suspension wires, calibration lines, and persistent narrow band
noise sources of unknown origin.

Time-frequency spectrograms are then created from conditioned data
for individual detectors from a series of Blackman-windowed discrete
Fourier transforms, of time length $\delta t$ set by the target
signal duration. A \emph{tile} is an estimate of the short-time
Fourier transform of the data at a specific time and frequency. Each
column in the tiling corresponds to a time bin of width $\delta t$
and each row corresponds to a frequency bin of width $\delta f$,
both linearly spaced, with $\delta f \delta t = 1$.  Adjacent time
bins overlap by $0.9 \delta t$ to guard against mismatch between
prospective signals and tiling time bins. Larger overlaps require
more computation and do not noticeably improve sensitivity\,\cite{kalmus08}.

In a one-detector search, we then have a complex-valued
time-frequency tiling from which we calculate the real-valued
one-sided PSD for every time bin.  To do this we multiply each tile
value by its complex conjugate and normalize the result to account
for sampling frequency and windowing function. We discard frequency
bins outside of the chosen search band.

In a two-detector search, we have two complex time-frequency tilings
(one for each detector) from which we calculate
 \be
   P^{(12)}_{tf} = \mathrm{Re}\left[T^{(1)}_{tf} \cc{T^{(2)}_{tf}} e^{-i2\pi f \Delta t}\right]
 \ee
where $T$ represents a tiling matrix and $t$ and $f$ are time and
frequency bin indices, and $(1)$ and $(2)$ denote the detector. Here
$\Delta t$ is the gravitational wave crossing time difference
between detectors;  this term takes care of applying the appropriate
time difference between detector data streams in the Fourier space,
with the advantage of permitting sub-sample time delays, which
significantly increases the sensitivity at higher frequencies. The
real part is kept, and normalization is applied as in the
one-detector case. To obtain a positive-definite statistic we take
the absolute value of each tile; this allows sensitivity to both
strongly correlated and strongly anti-correlated signals in the two
(potentially misaligned) detectors.

Next, we use off-source data to remove the background noise power
from each element of the PSD time-frequency tiling.  The elements
are fit to a gamma distribution\,\cite{kalmus08}, and outliers above a threshold
(typically four standard deviations) are discarded.  The resulting estimate on the mean is subtracted from each element
of the corresponding frequency bin in the PSD matrix, giving a
matrix of excess power (or ``cross excess power'' in the
two-detector case).  

To create analysis events we use a simple 2-dimensional clustering algorithm, allowing retention of signal energy which might otherwise be
fragmented in the case of extended signals in the time-frequency
plane.   Analysis events correspond to discrete clusters found by
the algorithm, and include information on cluster central frequency,
central time, bandwidth, duration, and so forth. The detection 
statistic (event loudness) is the sum over the cluster of tile
significance.

\subsection{T-Stack-a-flare pipeline}

The T-Stack pipeline combines burst events in the time domain.  

For each of $N$ burst events a trigger time is determined.  For a given GW detector,  $N$ time series segments containing those 
trigger times are then aligned to the trigger times, weighted according to antenna factor, and added together.  The 
resulting summed time series (either one or two, depending on how many detectors are included in the search) are then fed to 
the Flare pipeline.

As will be described below, the T-Stack pipeline has the advantage of achieving higher sensitivity in gaussian noise, but 
the disadvantage of being extremely sensitive to timing inaccuracies and variations in GW signal waveform from burst to burst.  

This makes it a potentially viable choice for analyzing multi-episodic events --- in which a single contiguous 100\,$
\mu$s-binned light curve might provide adequate timing precision --- but a poor choice for analyzing isolated burst 
events or incoherent signals such as band-limited WNBs. 

\begin{figure*}[!t]
\begin{center}
\includegraphics[angle=0,width=140mm, clip=false]{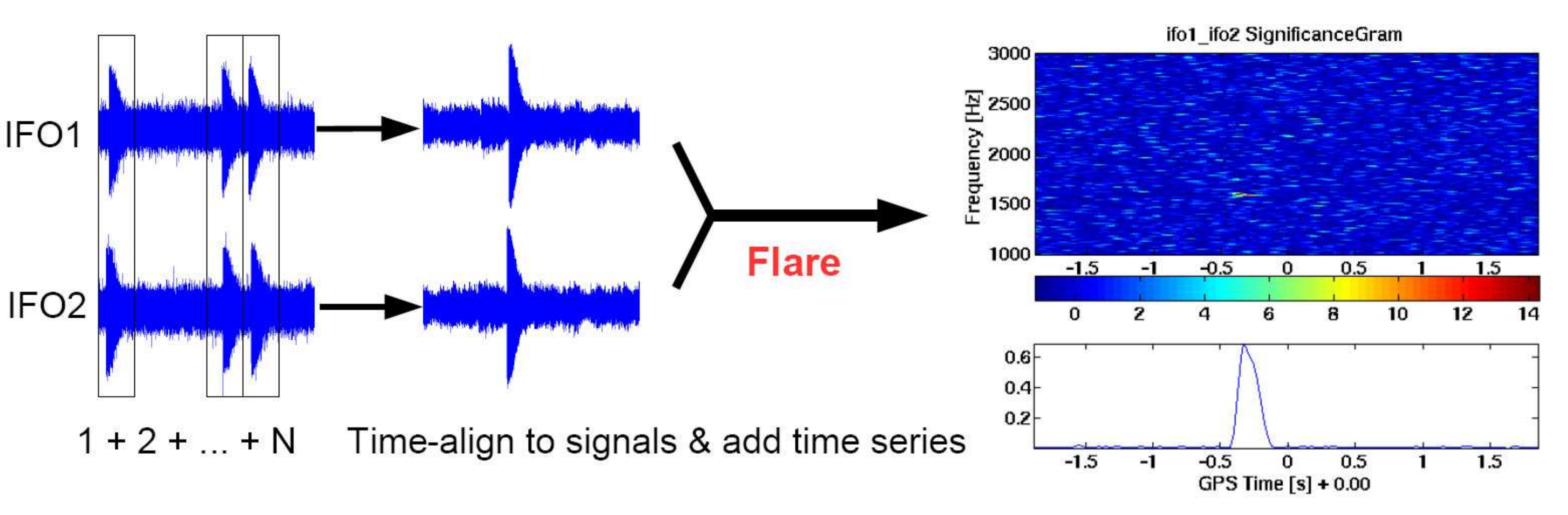}
\caption[Diagram of T-Stack pipeline] { Diagram of the T-Stack version of the Stack-a-flare pipeline.  The T-Stack 
pipeline has a thin layer added \emph{before} the  Flare pipeline in which gravitational wave data time series 
containing SGR burst event triggers are aligned on the trigger times and added together. These stacked time series 
are made for each detector and then run through the Flare pipeline as normal.}
\label{fig:tstackFlow}
\end{center}
\end{figure*}

\subsection{P-Stack-a-flare pipeline}

The P-Stack pipeline combines burst events in the frequency domain.  

We determine a trigger time for each of $N$ burst events based on gamma-ray data.  Each of $N$ time series containing those triggers is 
processed with the Flare pipeline, up to the clustering algorithm, exactly as in an individual SGR burst search. 
Antenna factors are applied at this time. The result is $N$ time-frequency significance tilings, with different noise but hopefully with similar or identical signals. The $N$ significance 
tilings are then aligned to the trigger time and added together.  The combined significance tiling is then fed through 
the Flare pipeline clustering algorithm with a fixed fraction of tiles to include in the clustering (e.g. 0.1\%).  A fixed 
fraction of tiles is used instead of a fixed loudness threshold value because  the variance of the tile loudness 
distribution at a given frequency increases with $N$.  

As will be described below, the P-Stack pipeline has the advantage of being relatively insensitive to timing 
inaccuracies or differences in waveform from burst to burst, but it has less sensitivity than the T-Stack pipeline for the 
(possibly unrealistic) precisely-known timing case, with deterministic waveforms. 

\begin{figure*}[!t]
\begin{center}
\includegraphics[angle=0,width=140mm, clip=false]{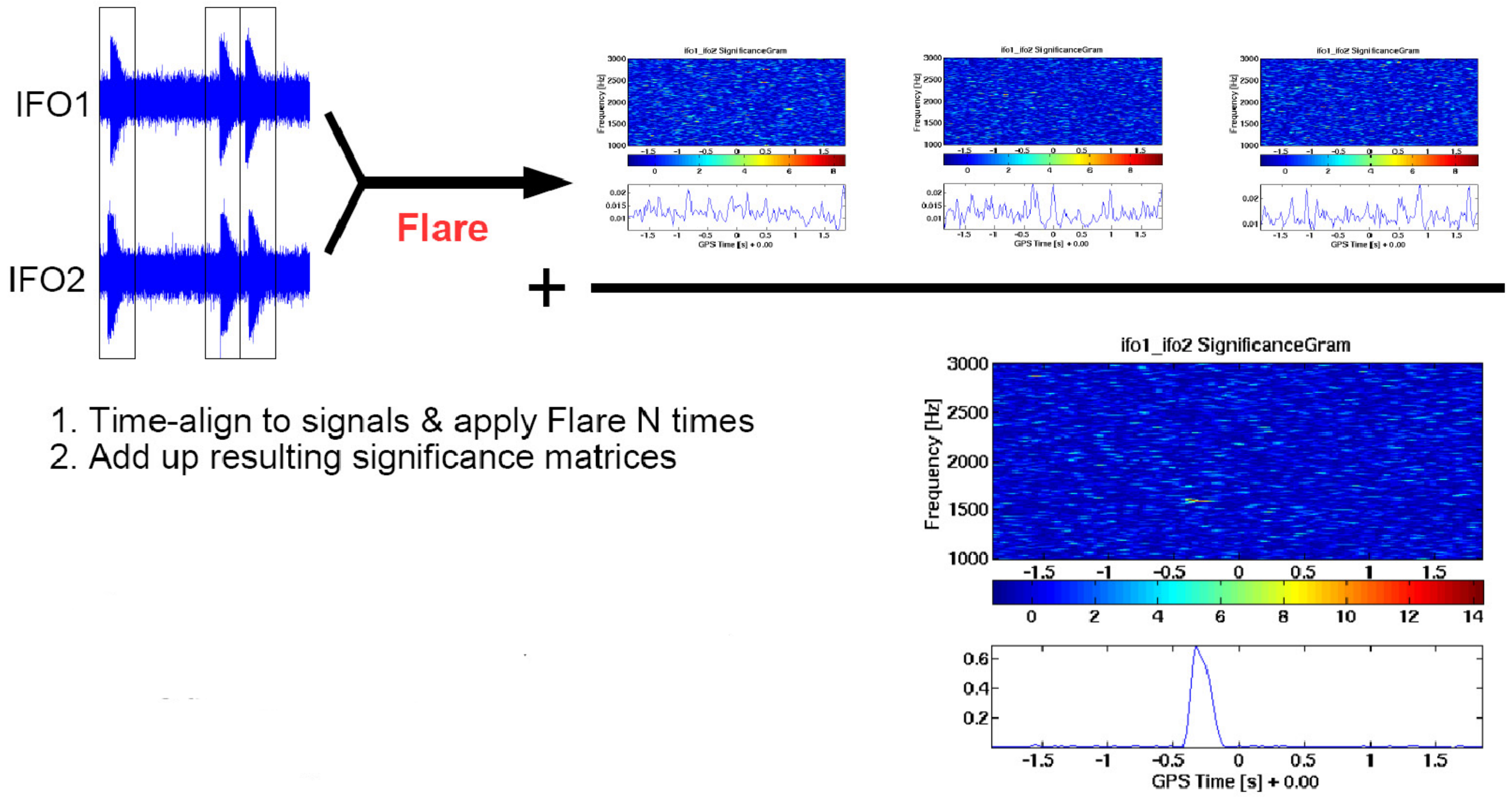}
\caption[Diagram of P-Stack pipeline] { Diagram of the P-Stack version of the Stack-a-flare pipeline.  The P-Stack 
pipeline has a thin layer added \emph{after} the  Flare pipeline in which gravitational wave data significance tilings 
containing SGR burst event triggers are aligned on the trigger times and added together.  Stacked significance tilings 
can then be run through the Flare pipeline clustering algorithm.
} \label{fig:fstackFlow}
\end{center}
\end{figure*}

\subsection{Loudest event upper limits}

\begin{figure}[t!]
\begin{center}
\includegraphics[angle=0,width=80mm, clip=false]{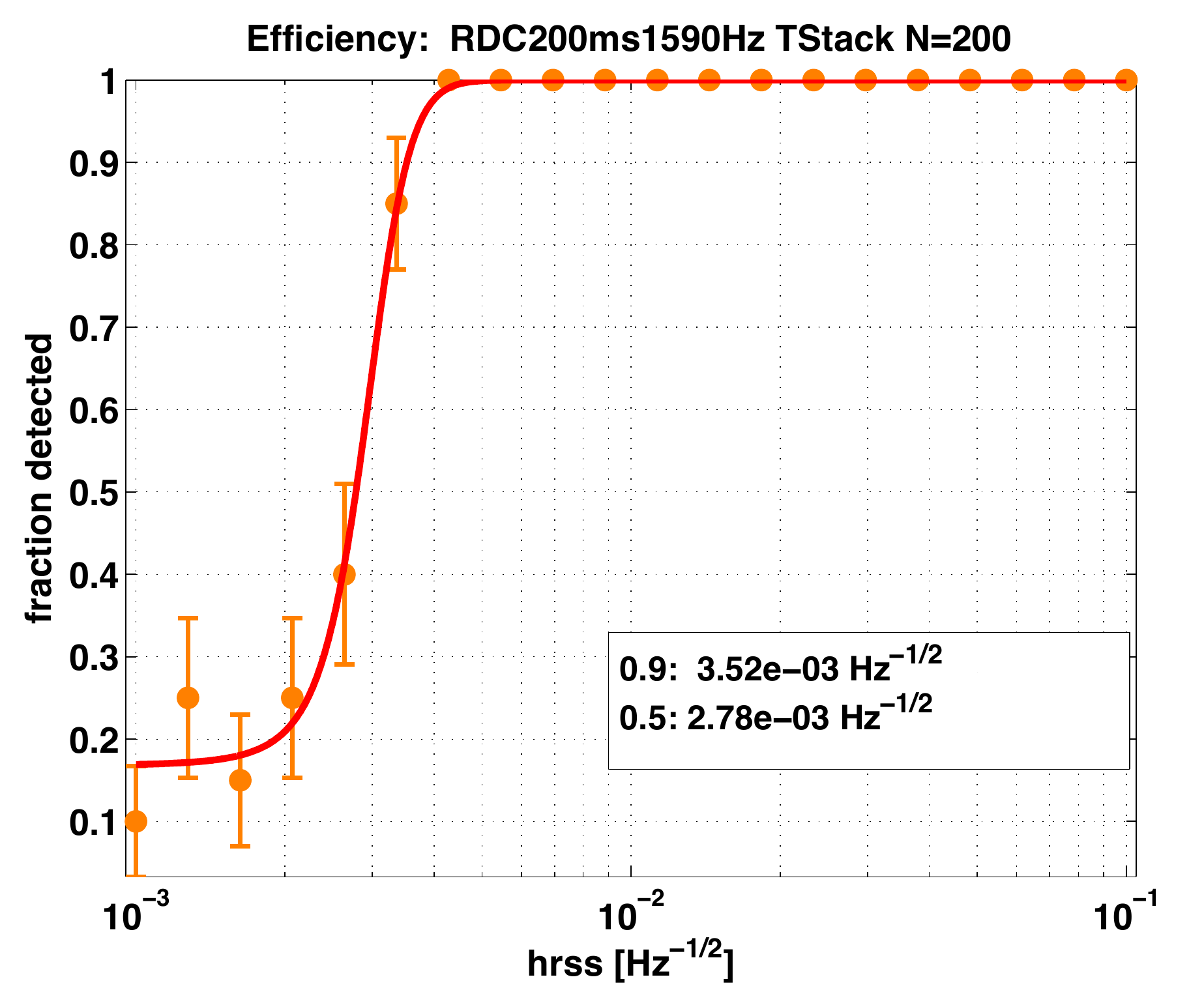}
\caption[Example efficiency curve  for Stack-a-flare sensitivity vs. $N$] { Example efficiency curve generated for the 
Monte Carlo experiment investigating Stack-a-flare sensitivity vs. $N$. This example curve was made with the T-Stack 
pipeline using compound injections of $N=200$ 1590\,Hz circularly polarized ringdowns injected into gaussian noise with 
$\sigma=1$.   Each efficiency curve was constructed using 20 amplitude scaling factors and 20 trials at each $\hrss$ 
amplitude.  Binomial error bars are given by $\sqrt{r(1-r)/M}$ where $r$ is the fraction detected at a given injection 
amplitude and $M$ is the number of injections in the Monte Carlo. } \label{fig:stackOfNSample}
\end{center}
\end{figure}

As in the individual SGR search, in the absence of a detection we can estimate loudest event upper limits\,
\cite{brady04} on GW root-sum-squared strain $\hrss$ incident at the detector, and GW energy emitted isotropically 
from the source assuming a nominal source distance.  

We estimate loudest-event upper limits\,\citep{brady04} on GW root-sum-squared 
strain $\hrss$ incident at the detector.  We can 
construct simulations of impinging GW with a given $\hrss$.   Following \cite{s2burst}
 \be 
 h_{\mathrm{rss}}^{2} = h_{\mathrm{rss+}}^{2} + h_{\mathrm{rss\times}}^{2}, 
 \ee
 where e.g.
 \be
 h_{\mathrm{rss+}}^{2} = \int_{-\infty}^{\infty} \abs{h_{+}}^{2} dt 
 \ee
and $h_{\mathrm{+,\times}}(t)$ are the two GW polarizations. The relationship between the GW polarizations and the detector response $h(t)$ 
to an impinging GW from a polar angle and azimuth $(\theta,\phi)$ and with polarization angle $\psi$ is:
 \be
 h(t) = F_{+}(\theta, \phi, \psi) h_+(t)   +  F_{\times}(\theta, \phi, \psi)
 h_{\times}(t)
 \label{eq:hsim}
 \ee
where $F_{+}(\theta, \phi, \psi)$ and $F_{\times}(\theta, \phi, \psi)$ are the antenna functions for the source at $(\theta,\phi)$\,\citep{300years}.   
At the time of the storm, the polarization-independent RMS antenna response $(F_{+}^2 + F_{\times}^2)^{1/2}$, which indicates the average sensitivity to a given sky location, was 0.39 for LIGO Hanford observatory and 0.46 for the LIGO Livingston observatory.

We can also set upper limits on the emitted isotropic GW emission energy $\egw$ at a source distance $R$ associated with $h_{+}(t)$ and ${h}_{\times}(t)$ via\,\citep{shapiro83}
 \be 
 E_{\mathrm{GW}} = 4\pi R^2 \frac{c^3}{16 \pi G} \int_{-\infty}^{\infty}\left((\dot{h}_{+})^2 + (\dot{h}_{\times})^2\right) dt. 
 \ee

The upper limit is computed in a frequentist framework following the commonly used procedure of 
injecting simulated signals in the background data and recovering them using the search pipeline\,(see for example 
\cite{S2inspiral,S2S3S4GRB}).  

An analysis event is associated with each injected simulation, and compared to the 
loudest on-source analysis event.  The GW strain or isotropic energy at e.g. 90\% detection efficiency is the strain or 
isotropic energy at which 90\% of injected simulations have associated events louder than the loudest on-source 
event.   

``Efficiency curves'' are constructed by the Flare pipeline by
repeatedly analyzing 4\,s background segments, each containing a single
simulation created with a range of $\hrss^{\mathrm{sim}}$ values,
and comparing the loudest simulation analysis event within 100\,ms
(for RDs) or 50\,ms (for WNBs) of the known injection time to the
loudest on-source analysis event\,\cite{kalmus08}.  An example efficiency curve is shown in Figure\,\ref{fig:stackOfNSample}.  The range of $\hrss$ values must
be chosen so that the smallest value produces simulations that are
always lost in the noise, and the largest value produces simulations
that are typically detected with large signal-to-noise ratios.  The
$\egw^{\mathrm{sim}}$ or $h_{\mathrm{rss}}^{\mathrm{sim}}$value at
90\% detection efficiency ($\egwn$ or $\hrssn$) occurs where 90\% of
the loudest simulation analysis events are larger than the loudest
on-source event.

For any given on-source region this results in four arrays of
numbers, each of which has length equal to the number of injected
simulations used to estimate the upper limit.  The first contains
the $\hrss$ values of injected simulations.  The second contains the
calculated $\egw$ values of injected simulations, or $\egw/R^2$ if
the distance to the source $R$ is not known.  The third contains the
loudness of the analysis event associated with the injected
simulation.  The fourth contains boolean values indicating whether
the associated analysis event was larger then the loudest on-source
analysis event or not.

The $\hrss$ and boolean (or the $\egw$ and boolean) arrays can be
used to construct the efficiency curve, with the $\hrss$ (or $\egw$)
values on the x-axis.  The y-axis indicates the fraction of analysis
events associated with an injected simulation of $\hrss$ as given by
the x-axis which are louder than the loudest on-source event.  In
the case of simulation $\hrss$ values which range over a discrete
set of scale factors, the y-axis value \emph{is} simply this
fraction. Binomial error bars may be added to these data points.

However, we are typically interested in the $\hrssn$ or $\hrssf$
value, that is the $\hrss$ of simulations whose associated analysis
event is louder than the loudest on-source analysis event 90\% or
50\% of the time.  Since we don't know this value ahead of time, it
is necessary to interpolate between the $\hrss$ values associated
with the discrete scale factors; we do this by fitting with
a sigmoid function.  The Flare efficiency curve fitting routine uses
two functions to perform these fits:  a four-parameter fit based on
the logistics function, and a five-parameter fit based on the
complementary error function.  The models were chosen
on empirical grounds, and details are given in \,\cite{kalmus08}.

We can follow the same procedure for the multiple burst search.  The only difference is the need to measure the $\hrss
$ or $\egw$ of a compound injection, instead of a simple (single) injection.

\subsection{Sensitivity dependence on $N$}
\label{section:sensitivityDependenceOnN}

The matched filter amplitude signal-to-noise ratio (SNR) is defined in the frequency domain as\,\cite{S1inspiral}
 \be
    \rho = \left[ 4 \int_0^\infty \  \frac{\tilde{h}(f)^2}{S_n (f)} df \right]^{1/2},
 \ee
where $\tilde{h}(f)$ is the Fourier transform of the signal time series and $S_n (f)$ is the noise power spectral density.  
Here, the numerator is the square root of the power in the signal.  In gaussian noise with zero mean, $S_n (f) = \sigma^2$, 
a constant. Since the standard deviation $\sigma$ of gaussian noise goes as the square root of $N$ and the amplitude of 
identical stacked signals goes as $N$, we expect the SNR of the optimal T-Stack algorithm for the recovery of identical 
signals from noise to go as $N^{1/2}$.

Whereas the T-Stack pipeline stacks amplitude, the P-Stack pipeline stacks power.  The background tiles in the power 
tiling at each individual frequency bin can be modeled as Gamma-distributed noise\,\cite{kalmus08}, for which the 
variance also goes as $N$, so we expect the power signal-to-noise ratio to increase as $N^{1/2}$. Since the 
amplitude goes as the square root of the power, we expect the P-Stack amplitude sensitivity to increase as $N^{1/4}$.

\begin{figure}[!t]
\begin{center}
\includegraphics[angle=0,width=80mm,clip=false]{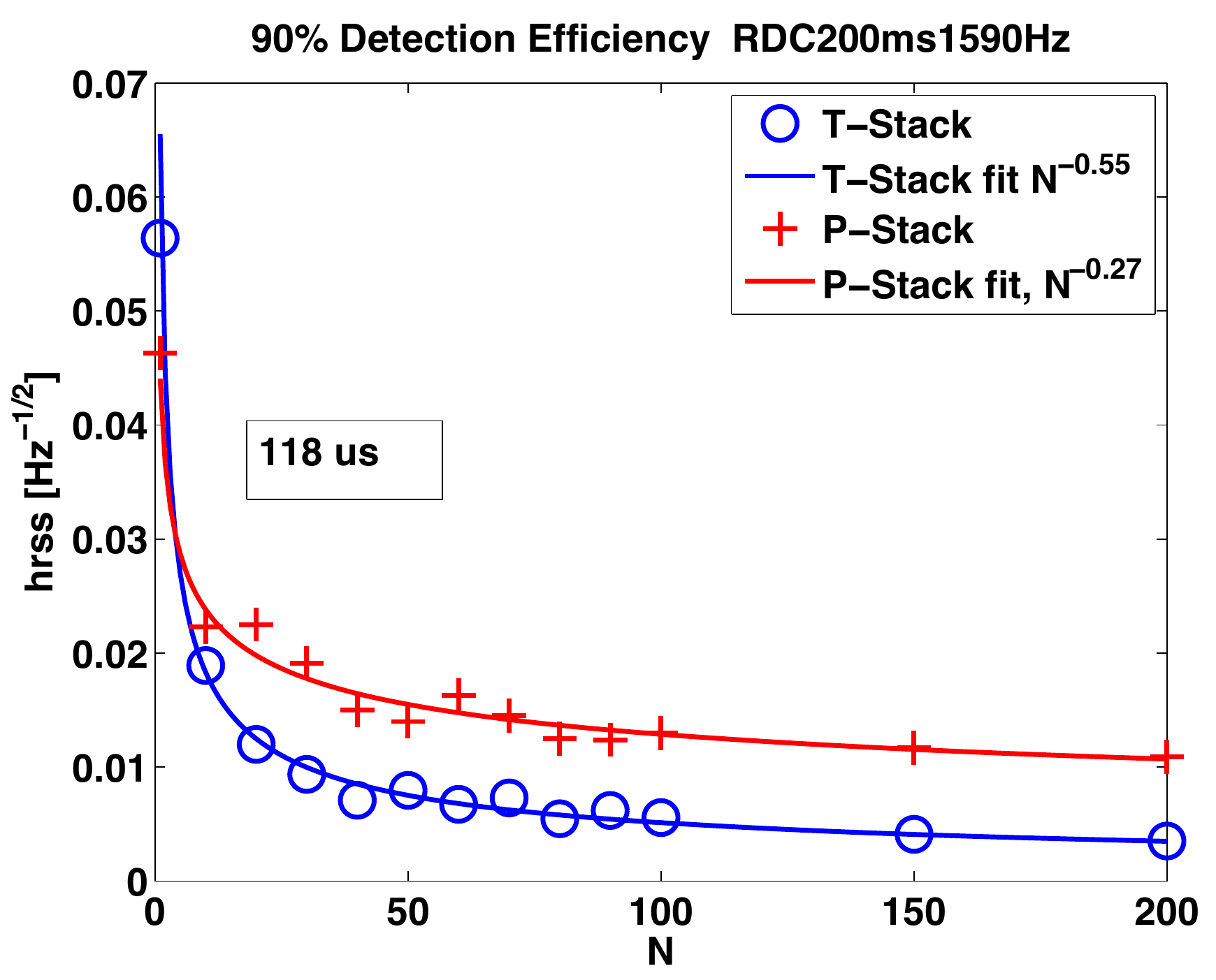}
\caption[T-Stack and P-Stack ringdown sensitivity dependence on $N$]
{T-Stack and P-Stack sensitivity dependence on $N$ at 90\%  detection efficiency, for 1590\,Hz $\tau$=200\,ms 
ringdowns in gaussian noise with $\sigma=1$.   The T-Stack pipeline show a sensitivity dependence of nearly $N^{1/2}$ 
and the results for the P-Stack pipeline show a sensitivity dependence of nearly $N^{1/4}$. All fits excluded the point 
$N=1$.  Results at 50\% detection efficiency look similar.}
\label{fig:ndepend}
\end{center}
\end{figure}

We tested these predictions by  injecting $N$ stacked 1590\,Hz 200\,ms $\tau$ ringdowns into gaussian noise with $
\sigma=1$. We then constructed efficiency curves  determining the injection $\hrss$ at 50\% and 90\% detection 
efficiency.  Each efficiency curve was constructed using 20 amplitude scaling factors and 20 trials at each $\hrss$ 
amplitude.  An example efficiency curve is shown in Figure\,\ref{fig:stackOfNSample}.  We then fit the 50\% and 90\% 
detection efficiency level results as functions of $N$ to a two-parameter power law of the form $y=A N^B$.  The results 
for both the T-Stack and P-Stack pipelines are shown in Figure\,\ref{fig:ndepend} at 90\% detection efficiency.  The fit 
for the T-Stack pipeline gives a sensitivity dependence in amplitude at both detection efficiency levels of  nearly 
$N^{1/2}$ ($N^{0.49}$ and $N^{0.55}$ for 50\% and 90\% detection efficiencies respectively), confirming our 
prediction.  This corresponds to an improvement in \emph{energy} of a factor of $N$.  The fit for the P-Stack pipeline 
gives a sensitivity dependence in amplitude at both detection efficiency levels of nearly $N^{1/4}$ ($N^{0.24}$ and 
$N^{0.27}$ for 50\% and 90\% detection efficiencies respectively), confirming our prediction.  This corresponds to an 
improvement in \emph{energy} of a factor of $N^{1/2}$.

\begin{figure}[!t]
\begin{center}
\includegraphics[angle=0,width=80mm,clip=false]{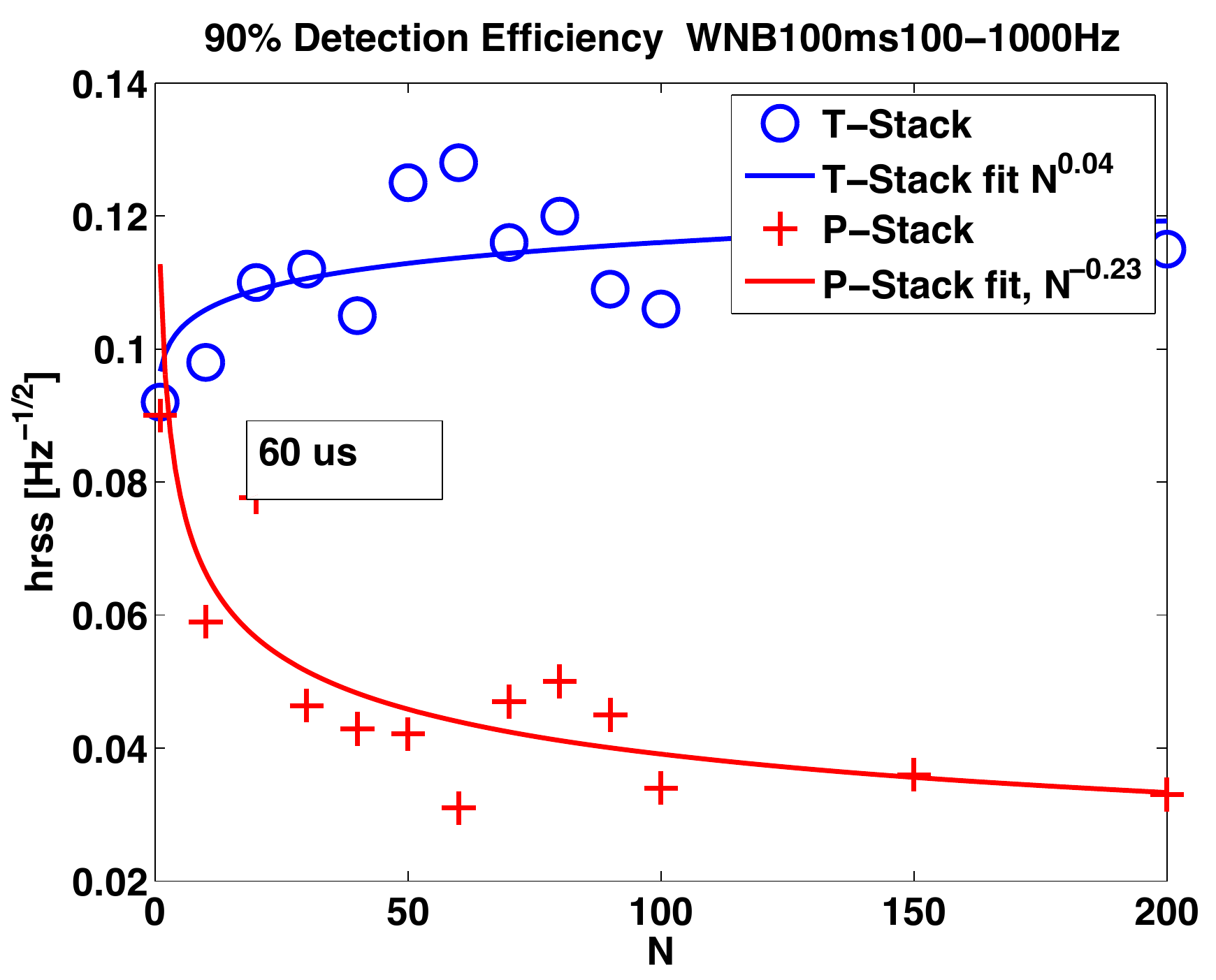}
\caption[T-Stack and P-Stack WNB sensitivity dependence on $N$] {T-Stack and P-Stack sensitivity dependence on 
$N$, at 90\% detection efficiency, for 100--1000\,Hz 100\,ms duration white noise bursts in gaussian noise with $
\sigma=1$. The results for the T-Stack pipeline show a sensitivity dependence of nearly $N^{0}$ (flat dependence), 
and the results for the P-Stack pipeline show a sensitivity dependence of nearly $N^{1/4}$, as in the coherent 
ringdown case. All fits exclude the point $N=1$.  Results at 50\% detection efficiency look similar.} 
\label{fig:ndependWNB}
\end{center}
\end{figure}

We repeated the experiment for 100\,ms duration 100--1000\,Hz band-limited WNBs.  In this case, we expected the 
coherent T-Stack pipeline to underperform the the P-Stack pipeline on these independently-generated stochastic 
incoherent signals. As expected, we found that the T-Stack pipeline shows no improvement as $N$ increases, while 
the P-Stack pipeline show the same $N^{1/4}$ sensitivity dependence seen in the coherent ringdown case 
($N^{0.36}$ and $N^{0.23}$ for 50\% and 90\% detection efficiencies respectively) .  The results are shown in Figure\,
\ref{fig:ndependWNB}; they illustrate the relative model-independence of the P-Stack pipeline.

\subsection{Sensitivity dependence on timing errors} \label{section:sensitivityVsTiming}

The T-Stack pipeline attains optimal sensitivity gains with increasing $N$ because it performs a phase coherent 
addition of signals.  We have shown that the P-Stack pipeline attains its $N^{1/2}$ energy sensitivity performance 
even in the case of stacked signals that are not coherent such as independently-generated white noise bursts.  For 
identical signals such as ringdowns, errors in the relative times between stacked signals cause breakdown of phase 
coherence.  

We performed Monte Carlo simulations using a simulated burst series with $N=20$ equal-amplitude ringdowns, and 
allowing the timing error to range. We characterized  1090\,Hz $\tau=200$\,ms and 2590\,Hz $\tau=200$\,ms circularly 
polarized ringdowns, corresponding to the low and high frequency ranges in the signal parameter space. Timing 
errors were randomly chosen for each ringdown from a normal distribution with $\sigma$ ranging from 10\,$\mu$s to 
100\,ms, and were applied as a timing shift to the given ringdown.  At each timing uncertainty we constructed 
efficiency curves using 20 amplitude scaling factors and 20 trials at each $\hrss$ amplitude.  

\begin{figure}[!t]
\begin{center}
\includegraphics[angle=0,width=80mm,clip=false]{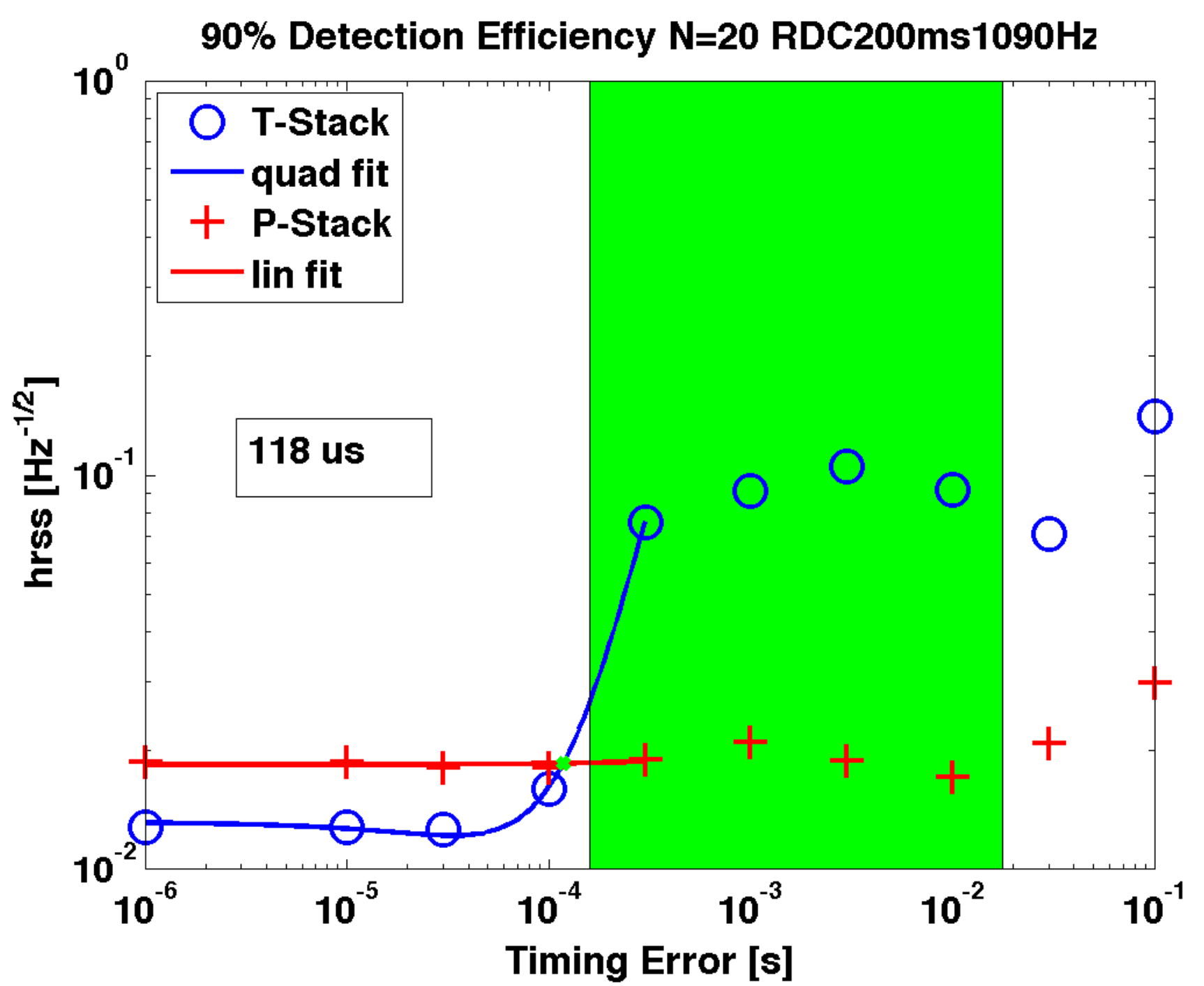}
\caption[T-Stack and P-Stack sensitivity vs. timing error, 1090\,Hz RDs] { T-Stack and P-Stack sensitivity versus timing 
error, for 1090\,Hz $\tau=200$\,ms circularly polarized RD,  $N=20$, for $\hrssn$.  T-Stack is more sensitive for small 
timing errors, but degrades.  The crossover point is noted;  for T-Stack to be effective at 1090\,Hz, apparently timing 
error must be $\lesssim100$\,$\mu$s at one-sigma.  T-Stack results level off at high timing errors (greater than $\sim 
\sci{2}{-4}$, or $\sim$90 degrees of phase) because the Monte Carlo effectively randomizes the phases of the stacked 
signals.  The green region shows the approximate range of timing uncertainties, shown in Table\,\ref{table:peaks}.  
Results at 50\% detection efficiency look similar.} \label{fig:stackOfE1090}
\end{center}
\end{figure}

\begin{figure}[!t]
\begin{center}
\includegraphics[angle=0,width=80mm,clip=false]{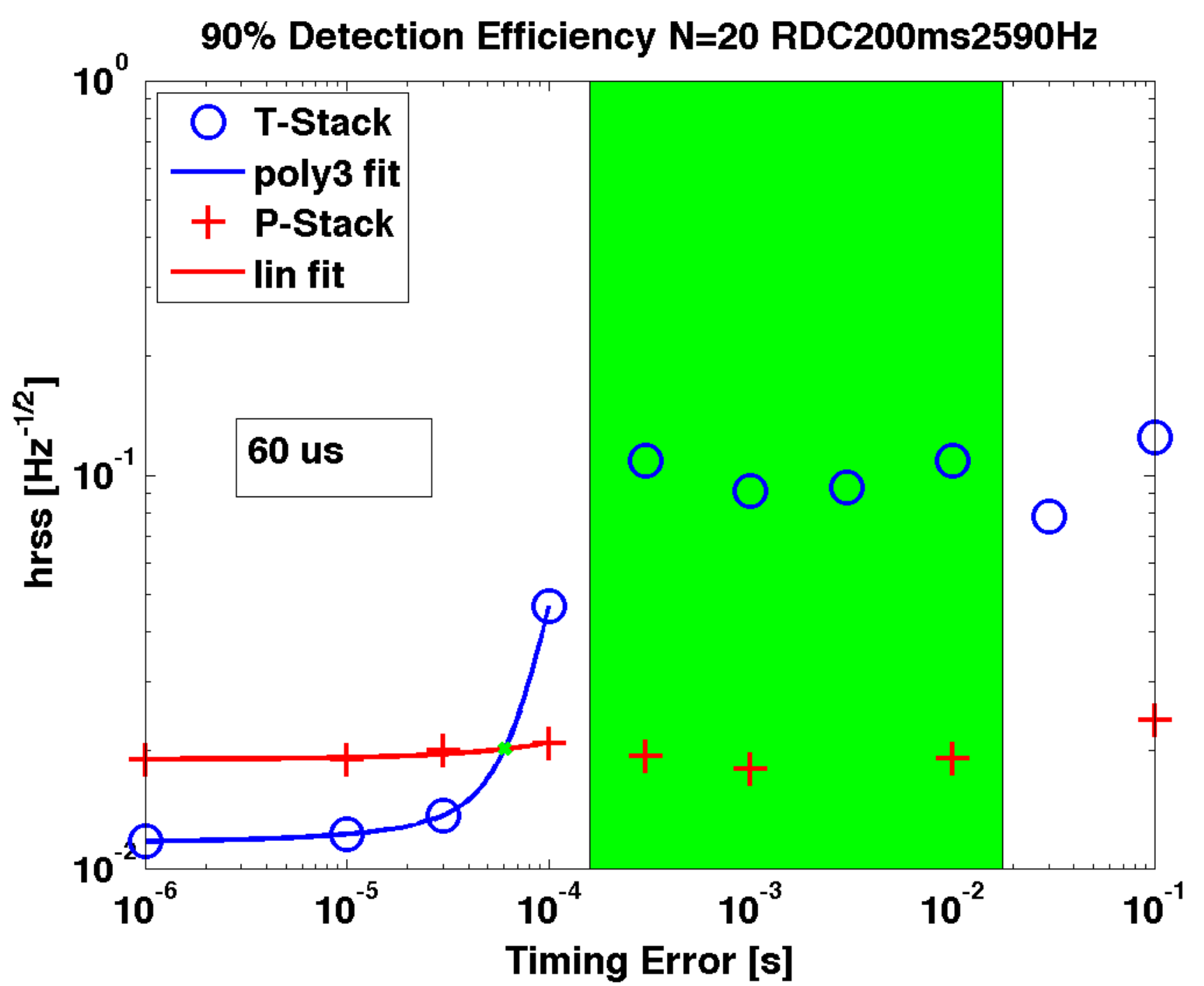}
\caption[T-Stack and P-Stack sensitivity vs. timing error, 2590\,Hz RDs] { T-Stack and P-Stack sensitivity versus timing 
error, for 2590\,Hz $\tau=200$\,ms circularly polarized RD,  $N=20$, for $\hrssn$. T-Stack is more sensitive for small 
timing errors, but degrades.  The crossover point is noted;  for T-Stack to be effective at 2590\,Hz, apparently, timing 
error must be $\lesssim50$\,$\mu$s at one-sigma.   T-Stack results level off at high timing errors (greater than $\sim 
\sci{1}{-4}$, or $\sim$90 degrees of phase) because the Monte Carlo effectively randomizes the phases of the stacked 
signals.  The green region shows the approximate range of timing uncertainties, shown in Table\,\ref{table:peaks}.  
Results at 50\% detection efficiency look similar.} \label{fig:stackOfE2590}
\end{center}
\end{figure}

The results are displayed in Figure\,\ref{fig:stackOfE1090} (1090\,Hz ringdowns) and Figure\,\ref{fig:stackOfE2590} 
(2590\,Hz ringdowns) at 90\% detection efficiency.  As expected, the P-Stack method is independent of timing error, up 
until large timing errors on the order of the signal duration. The T-Stack pipeline, on the other hand, shows a 
pronounced dependence on timing error, especially in the case of high frequency simulations.  Each plot shows data 
for both T-Stack and P-Stack pipelines, and finds the equal-sensitivity timing error (P-Stack and T-Stack curve 
intersection point) using polynomial fits.

For the T-Stack pipeline to be effective at 1090\,Hz, apparently, timing error must be $\lesssim100$\,$\mu$s at one-
sigma.  For the T-Stack pipeline to be effective at 2590\,Hz, timing error must be $\lesssim50$\,$\mu$s at one-sigma.  
For the $N=20$ case shown, with no timing error the T-Stack pipeline is about 1.5 times more sensitive than the
P-Stack pipeline.  

\subsection{Implications of the pipeline characterization}

We summarize the implications from characterizing the T-Stack and P-Stack pipelines.  We envision four possible 
types of stacked SGR searches:
 \ben
 \i High frequency (1000--3000\,Hz) searches for ringdown burst  emission,  for single SGR storm events (ringdown 
upper limits);
 \i Low frequency (100--1000\,Hz) searches for stochastic burst  emission,  for single SGR storm events (band- and 
time-limited WNB upper limits);
 \i High frequency (1000--3000\,Hz) searches for ringdown burst  emission, for isolated, time-separated SGR bursts 
(ringdown upper limits);
 \i Low frequency (100--1000\,Hz) searches for stochastic burst emission,  for isolated, time-separated SGR bursts 
(band- and time-limited WNB upper limits).
\een

We distinguish storm events from isolated events because it is much easier to get precise relative times for storm 
events.  

We have found that the P-Stack pipeline should be effective in any of these cases, with an energy sensitivity gain over 
the individual burst search of approximately $N^{1/2}$. The T-Stack pipeline shows an energy sensitivity improvement 
of approximately $N$, but only if the target signals are coherent, and only if the relative timing between SGR GW burst 
events can be known to high precision ($\lesssim$100\,$\mu$s at 1090\,Hz and $\lesssim$50\,$\mu$s at 2590\,Hz).  
As we will see, this timing precision requirement is stringent.   In the future we might search for a method of overcoming the T-Stack timing precision limitation, possibly by using additional computational resources to perform time shifts between the time series in the stack.

\section{ SGR 1900+14 storm mock search} \label{section:stacksim}

In this section we describe a mock stacking search with the P-Stack pipeline, using simulated LIGO data, for GW 
associated with the 2006 March 29 SGR 1900+14 storm event.  We first describe a procedure for estimating burst start 
times and fluences from the light curve.  We then discuss GW emission models.  We finally present sensitivity estimates.

\subsection{Light curve and GW emission models}
\label{section:light curve}

Data from the BAT detector on the Swift satellite are publicly available.  In Figure\,\ref{fig:lc_raw}, we show the storm 
light curve in photon counts in the 15--100\,keV band with 1\,ms time bins.  SGR bursts lasting longer than 500\,ms are 
typically considered ``intermediate flares'';  the storm contained both intermediate flares and common bursts.  

Before choosing GW emission models for stacking, we gathered information from the light curve, most importantly consistent relative 
burst times.  Though the BAT timing resolution is 100\,$\mu$s, additional resolution would not improve our estimates 
of relative burst times as we shall see.    We also measured integrated counts under each burst, a quantity roughly 
proportional to fluence.  These efforts were complicated by overlapping and non-uniform bursts in the noisy light 
curve.   We found that a 30-sample running average of the 1\,ms-binned light curve aided aspects of the analysis.

\begin{table}
\caption{The 18 most energetic peaks in the storm light curve included in the analysis, ordered by time.  First column is 
peak number.  Second column is the time in seconds relative to the beginning of the light curve shown (see Figure\,
\ref{fig:lc_raw}), with the one-sigma uncertainty from the rising edge fit given.  Third column is light travel time from 
satellite to geocenter in ms, which is applied before analysis.   Fourth column is integrated burst counts.  Fifth column 
is estimated burst fluence, based on a conservative estimate of $\sci{1.0}{-4}$\,erg\,cm$^{-2}$ for the total fluence  of 
the storm by the Konus-Wind team, in the 20--200\,keV range\,\cite{gcn4946}.   Sixth column is fitted rising edge slope 
in counts/s.  Seventh column is the approximate duration of the peak in seconds.}
\begin{tabular}{@{\extracolsep{\fill}}lrccccc}
 \hline \hline
no. & time & delay & counts & fluence                & rise  & dur. \\ 
                  & [s]    &  [ms]           &              &  [erg cm$^{-2}$]  &  [counts/s] &  [s] \\
 \hline 
 1  & 0.975  $\pm$ 0.001  & 17.48  & 1.1e+05  & 5.0e-06  & 1.64e+04  & 0.16  \\
 2  & 1.273  $\pm$ 0.002  & 17.46  & 2.9e+04  & 1.4e-06  & 3.30e+04  & 0.06  \\
 3  & 1.973  $\pm$ 0.003  & 17.46  & 5.0e+05  & 2.4e-05  & 1.58e+04  & 0.94  \\
 4  & 4.067  $\pm$ 0.003  & 17.43  & 2.3e+04  & 1.1e-06  & 1.51e+04  & 0.09  \\
 5  & 4.170  $\pm$ 0.006  & 17.43  & 7.9e+04  & 3.7e-06  & 1.08e+04  & 0.15  \\
 6  & 4.423  $\pm$ 0.018  & 17.43  & 2.6e+05  & 1.2e-05  & 2.51e+03  & 0.57  \\
 7  & 6.819  $\pm$ 0.003  & 17.41  & 9.1e+04  & 4.3e-06  & 1.61e+04  & 0.20  \\
 8  & 7.179  $\pm$ 0.001  & 17.40  & 2.4e+04  & 1.1e-06  & 2.10e+04  & 0.08  \\
 9  & 10.883  $\pm$ 0.002  & 17.36  & 1.1e+05  & 5.1e-06  & 2.28e+04  & 0.26  \\
 10  & 15.287  $\pm$ 0.004  & 17.30  & 2.2e+04  & 1.0e-06  & 9.58e+03  & 0.11  \\
 11  & 15.822  $\pm$ 0.003  & 17.30  & 1.8e+05  & 8.4e-06  & 7.45e+03  & 0.47  \\
 12  & 16.603  $\pm$ 0.004  & 17.29  & 3.7e+05  & 1.7e-05  & 1.60e+04  & 0.77  \\
 13  & 17.632  $\pm$ 0.002  & 17.28  & 3.2e+04  & 1.5e-06  & 2.81e+04  & 0.10  \\
 14  & 18.298  $\pm$ 0.009  & 17.27  & 4.7e+05  & 2.2e-05  & 1.08e+04  & 1.22  \\
 15  & 19.718  $\pm$ 0.000  & 17.25  & 3.1e+04  & 1.4e-06  & 2.78e+04  & 0.11  \\
 16  & 20.865  $\pm$ 0.013  & 17.24  & 1.7e+05  & 8.1e-06  & 3.60e+03  & 0.62  \\
 17  & 22.284  $\pm$ 0.011  & 17.22  & 2.9e+05  & 1.4e-05  & 9.43e+03  & 0.57  \\
 18  & 30.335  $\pm$ 0.001  & 17.12  & 3.4e+04  & 1.6e-06  & 1.65e+04  & 0.15  \\
 \hline \label{table:peaks} 
 \end{tabular} 
 \end{table}

\begin{figure*}[!t]
\subfigure{
\includegraphics[angle=0,width=180mm,clip=false]{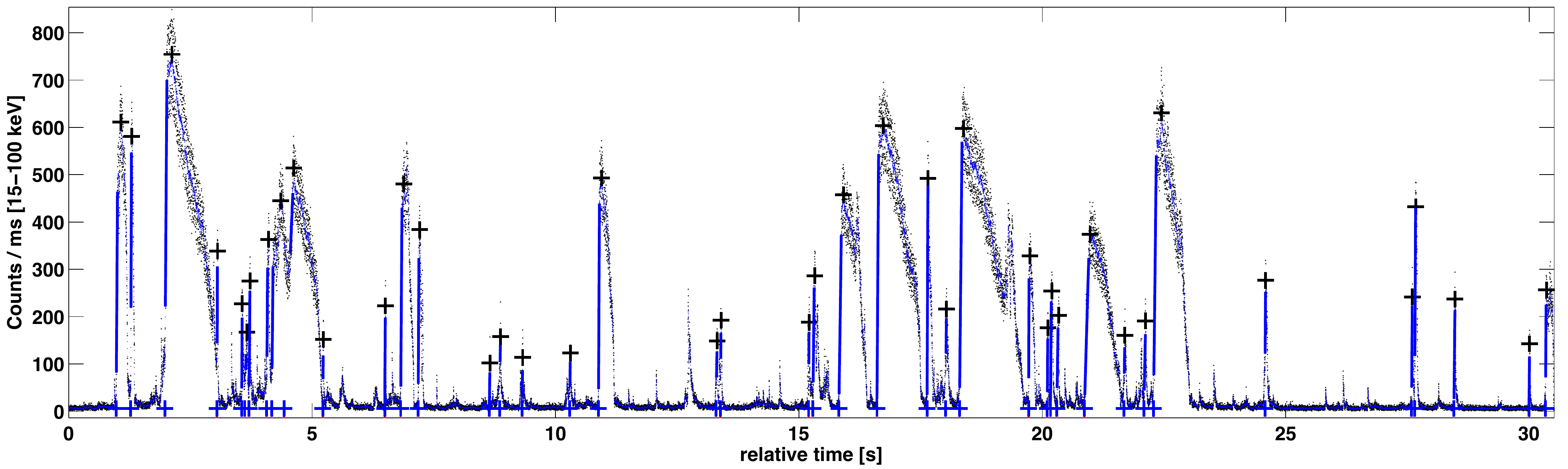}}
\subfigure{
\includegraphics[angle=0,width=180mm,clip=false]{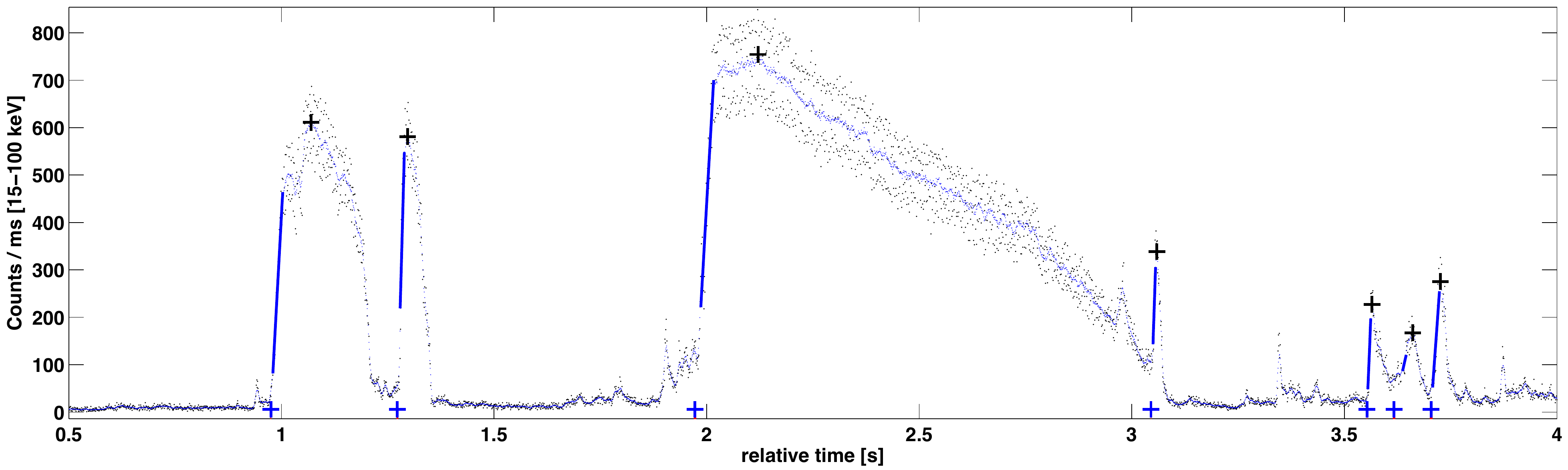}}
\caption[] { SGR 1900+14 storm light curve with 1\,ms bins in the (15--100)\,keV band.  Bottom plot shows a detail.  Burst start times are estimated by fitting the steeply rising burst edges; 
EM fluences are estimated  by integrating light curve area under each burst.   A 30-bin running average is shown in addition to the raw light curve.  Solid lines are linear fits to rising 
edges; the boundaries of rising edges were found by examining the first derivatives in the neighborhoods of the peak locations.  Crosses mark burst peaks and intersections of the rising edge fits extrapolated to a linear fit of the noise floor measured in a quiescent period of data in the 50\,s BAT sequence before the start of the storm.  The one-sigma timing uncertainty averaged over all measurements is $2.9$\,ms.    X-axis times are relative to 2006-03-29 02:53:09.9 UT at the Swift satellite.}
\label{fig:lc_rises}
\end{figure*}

\begin{figure}[!t]
\begin{center}
\includegraphics[angle=0,width=80mm, clip=false]{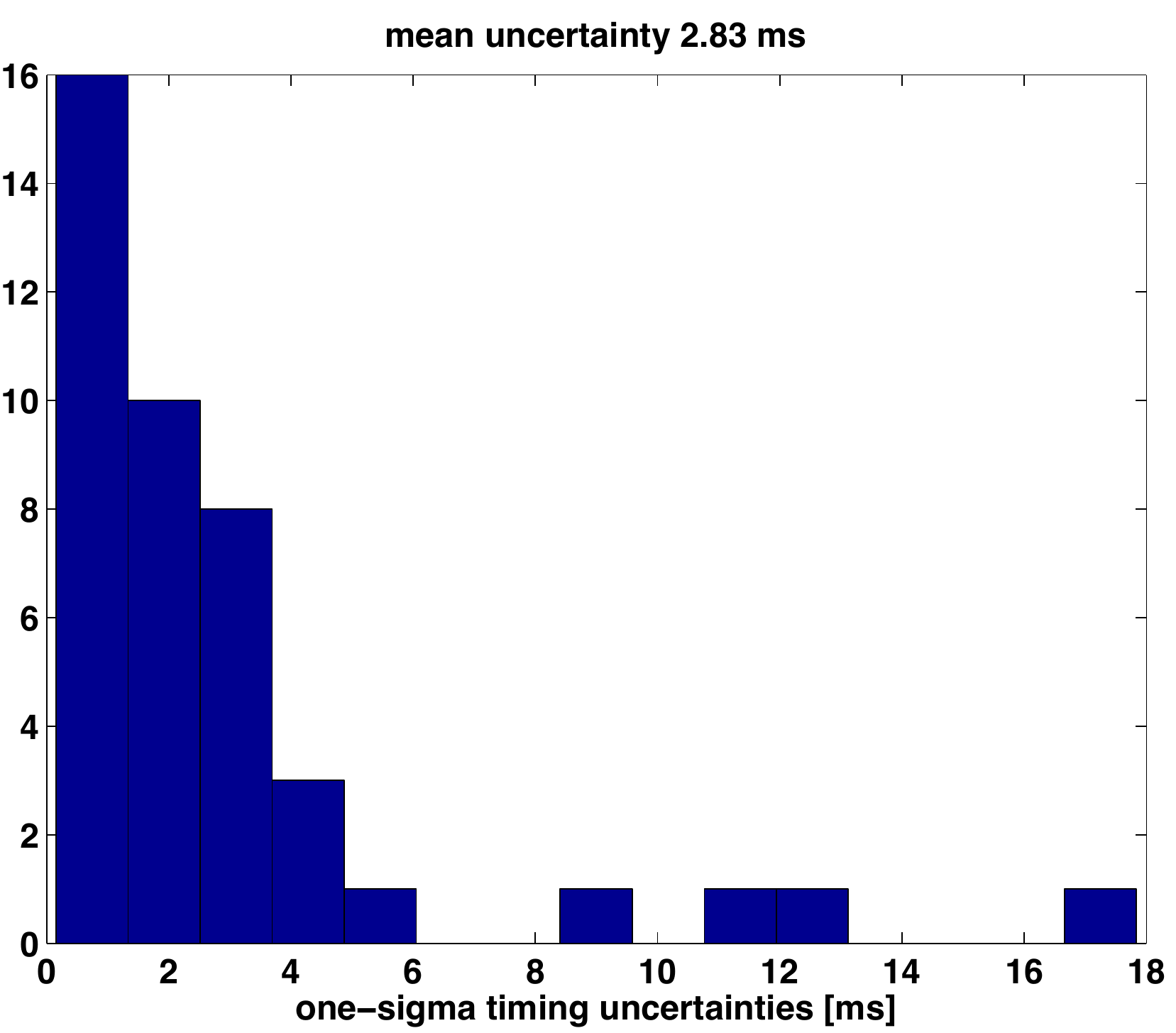}
\caption[] {Histogram of one-sigma timing uncertainties associated with the rise start time of each burst.  These 
uncertainties were estimated from linear fits of the rising edges of bursts in the light curve assuming errors in the light 
curve data are independent normal with constant variance.  The smallest uncertainty is $\sci{1.6}{-04}$\,s, the largest 
is 18\,ms, and the mean is 2.8\,ms.  The largest uncertainty is due to a burst at $\sim$4.4\,s which rises out of another 
large burst, far above the noise floor.  The other large uncertainties are due to bursts with rippling rising edges which 
may be due to additional injections of energy.} \label{fig:timingUncert}
\end{center}
\end{figure}

We chose the point of intersection of the rapid rising edges of each burst with the light curve noise floor as plausible 
and consistent estimates of the times at which possible progenitor catastrophic neutron star events occurred.    So 
long as GW emission is shifted less than 1.5\,s before or after this time, it should be well within the on-source region.    
We first used a generic peak-finding routine to fit and find approximate peak locations; the smallest peaks were 
ignored as insignificant given the GW emission stacking models we ultimately chose.  Local maxima were found by exploring the 
neighborhoods around the fitted peaks.  Walking left from local maxima, a tuned first derivative threshold was used to 
determine the top and bottom of each burst's rising edge.  The rising edge between these bounds was then fit with a 
line.    Figure\,\ref{fig:lc_rises} shows the fits on all bursts considered for inclusion in the analysis.  We then corrected each burst time for the light travel time to the 
geocenter using the known SGR 1900+14 sky position and known satellite positions.  The light travel delay at the 
beginning of the light curve is 17.48\,ms (arriving at the satellite first) and at the end is 17.12\,ms, a change of 0.35\,ms 
over a course of 30.5\,s.   A histogram of one-sigma uncertainty estimates is shown in Figure\,\ref{fig:timingUncert}.  
These uncertainty estimates were applied individually to respective simulations in the Monte Carlo.

\begin{figure*}[!t]
\subfigure{
\includegraphics[angle=0,width=180mm,clip=false]{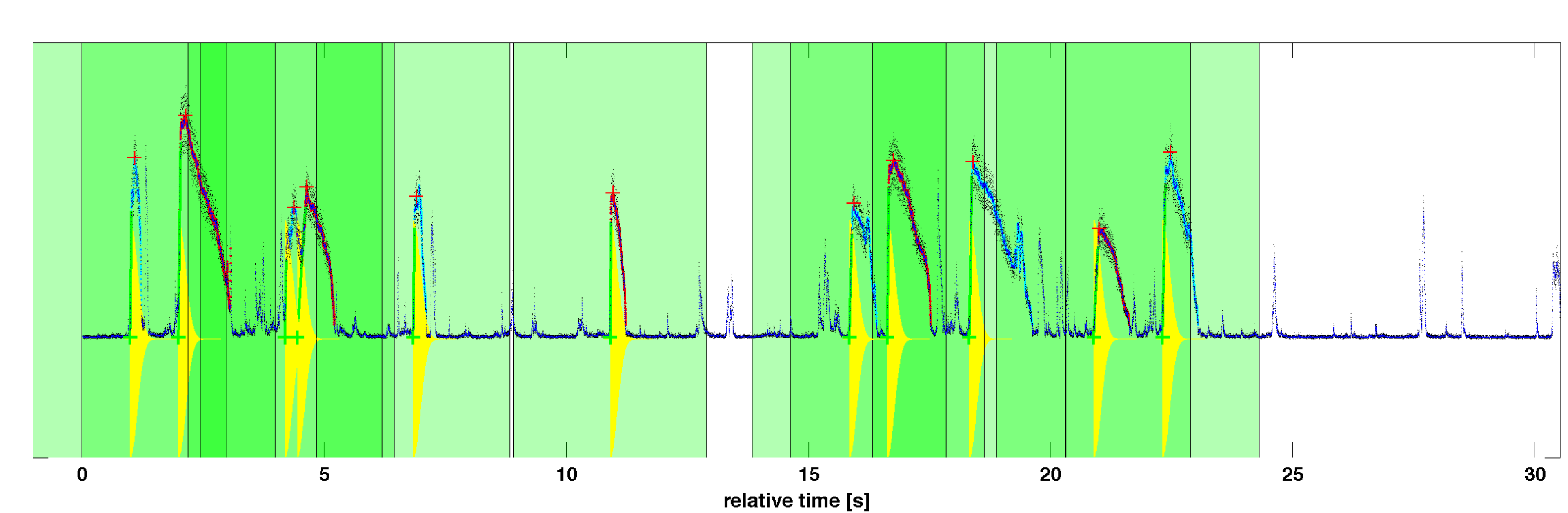}}
\subfigure{
\includegraphics[angle=0,width=180mm,clip=false]{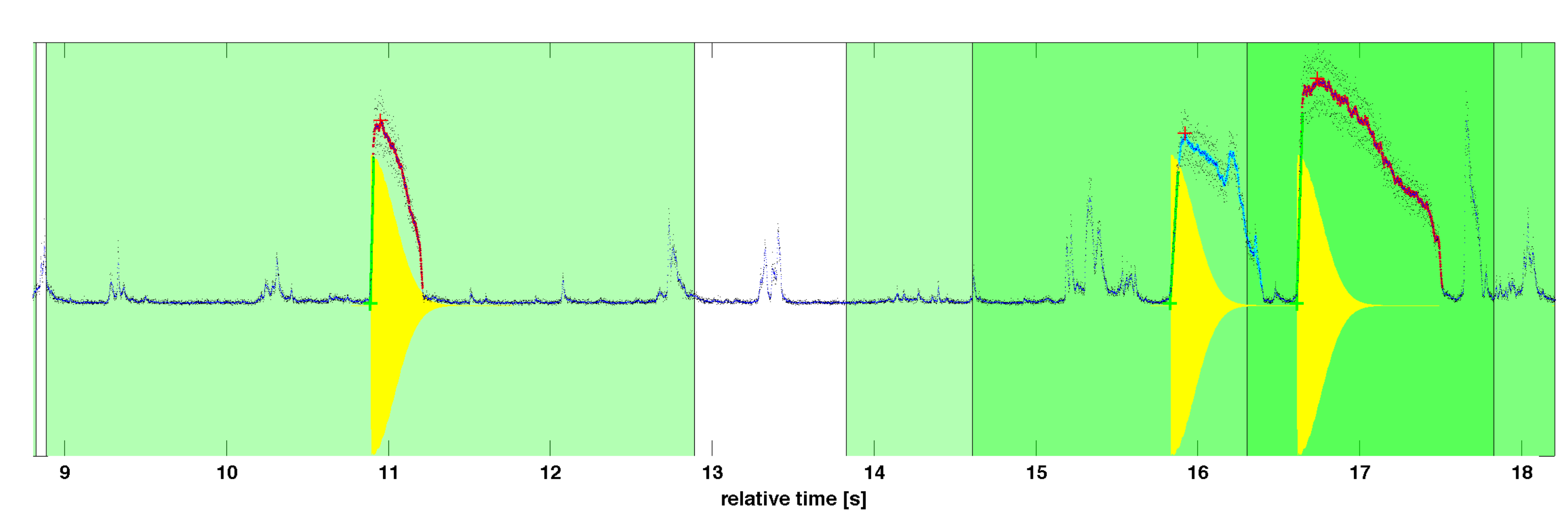}}
\caption[] { SGR 1900+14 storm light curve, focusing on fluence (integrated counts) estimates and timescales.  The 
$N=11$ flat model is illustrated here.  In addition to features common to Figure\,\ref{fig:lc_rises}, alternating red and 
cyan highlights indicate integration bounds for fluence estimation of the 11 most energetic bursts.  Fluence estimation 
is one of the goals of the light curve processing algorithm.  The yellow features are 1090\,Hz $\tau=200$\,ms ringdown 
waveforms,  aligned with the nominal beginning of the electromagnetic bursts (indicated as before by green circles).  
A small constant time offset between this nominal time and gravitational wave emission would be immaterial to the 
search, so long as the time offset were small relative to the on-source region half-duration of 2\,s.  The $\pm2$\,s on-
source regions, which often overlap even in this $N=11$ model, are illustrated by the green shading.  }
\label{fig:lc_timescales}
\end{figure*}

Integrated counts under each burst were also estimated, by walking to the right along the smoothed light curve from 
peak markers until nearly reaching the noise floor or encountering the beginning of the next burst's rising edge.   The 
light curve with boundaries used in the integrated counts estimate indicated is shown in Figure\,\ref{fig:lc_timescales}.

Burst integrated count uncertainties were conservatively though qualitatively estimated to be less than 5\%, except for 
one overlapping burst with a burst rise at about 4.2\,s.  Uncertainty in the overlapping peak is astrophysical in nature 
and no larger 
than 3\% of the total storm ßuence.  The Konus-Wind team reported a conservative storm total fluence of $\sci{1-2}
{-4}$\,erg\,cm$^{-2}$ in the 20--200\,keV range\,\cite{gcn4946}.   To convert integrated counts to fluences, we used the 
most conservative estimate of ${1.0}{-4}$\,erg\,cm$^{-2}$ and the total integrated counts in the entire storm to obtain 
fluence per count and thus fluences for each burst in the storm.  In doing so we assume bursts exhibit spectral 
uniformity.  We wish to know the absolute fluence in order to set upper limits on $\gamma \equiv \egwn / \eem$.  
Uncertainty in $\gamma$ from integrated counts estimates is negligible compared to the uncertainty from the 
conversion, and so we simply absorb it there.  Burst measurements are given in Table\,\ref{table:peaks}.

\begin{figure}[!t]
\begin{center}
\includegraphics[angle=0,width=80mm, clip=false]{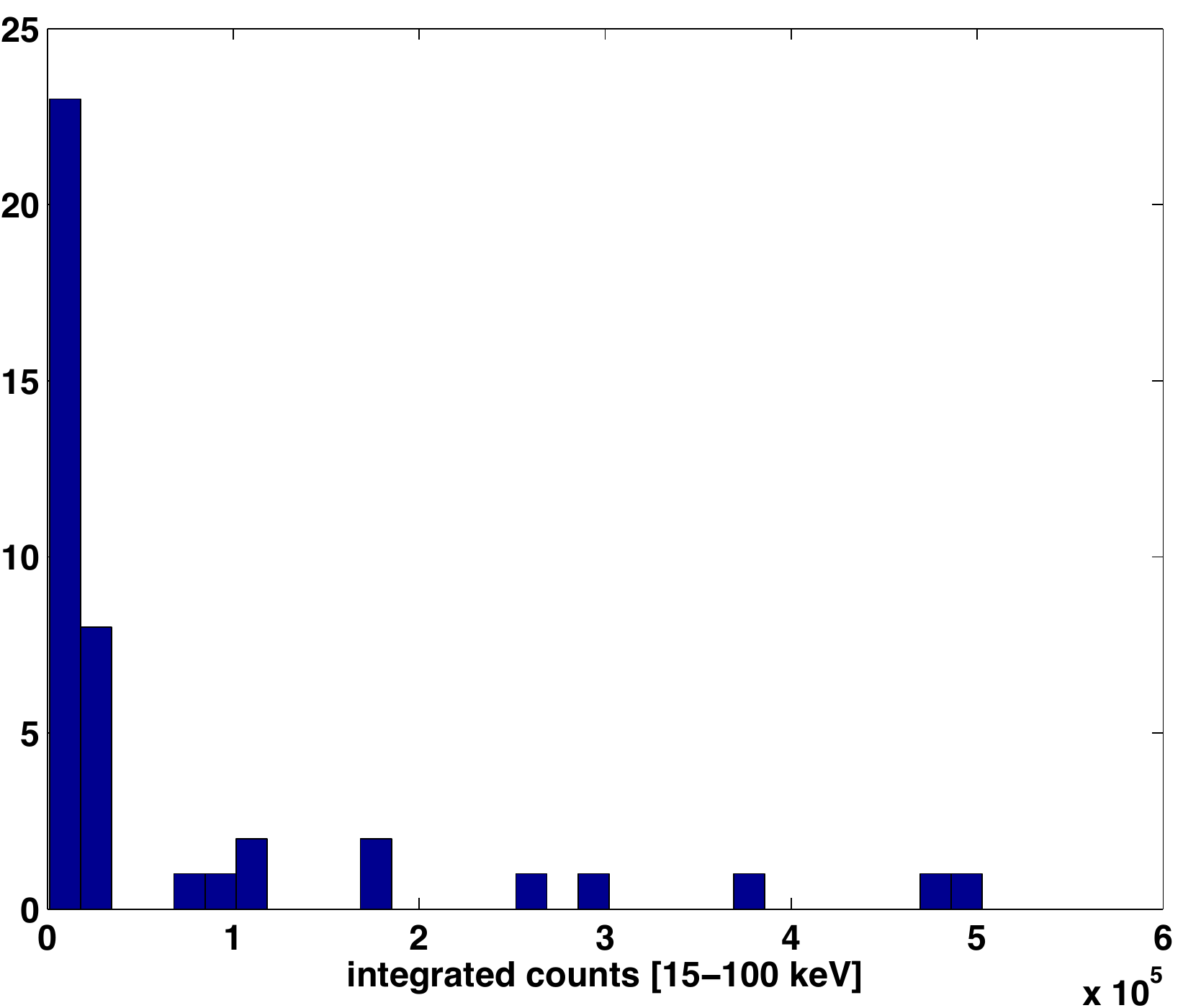}
\caption[Histogram of integrated counts of bursts in the SGR 1900+14 storm.] { Histogram of integrated counts in 
bursts in the SGR 1900+14 storm.  This quantity is roughly proportional to fluence. The histogram shows a separation 
between the 11 most electromagnetically energetic bursts and the rest.  This separation determined the break point for 
the $N=11$ flat GW emission model. } \label{fig:fluenceHist}
\end{center}
\end{figure}

\begin{figure}[!t]
\begin{center}
\includegraphics[angle=0,width=80mm, clip=false]{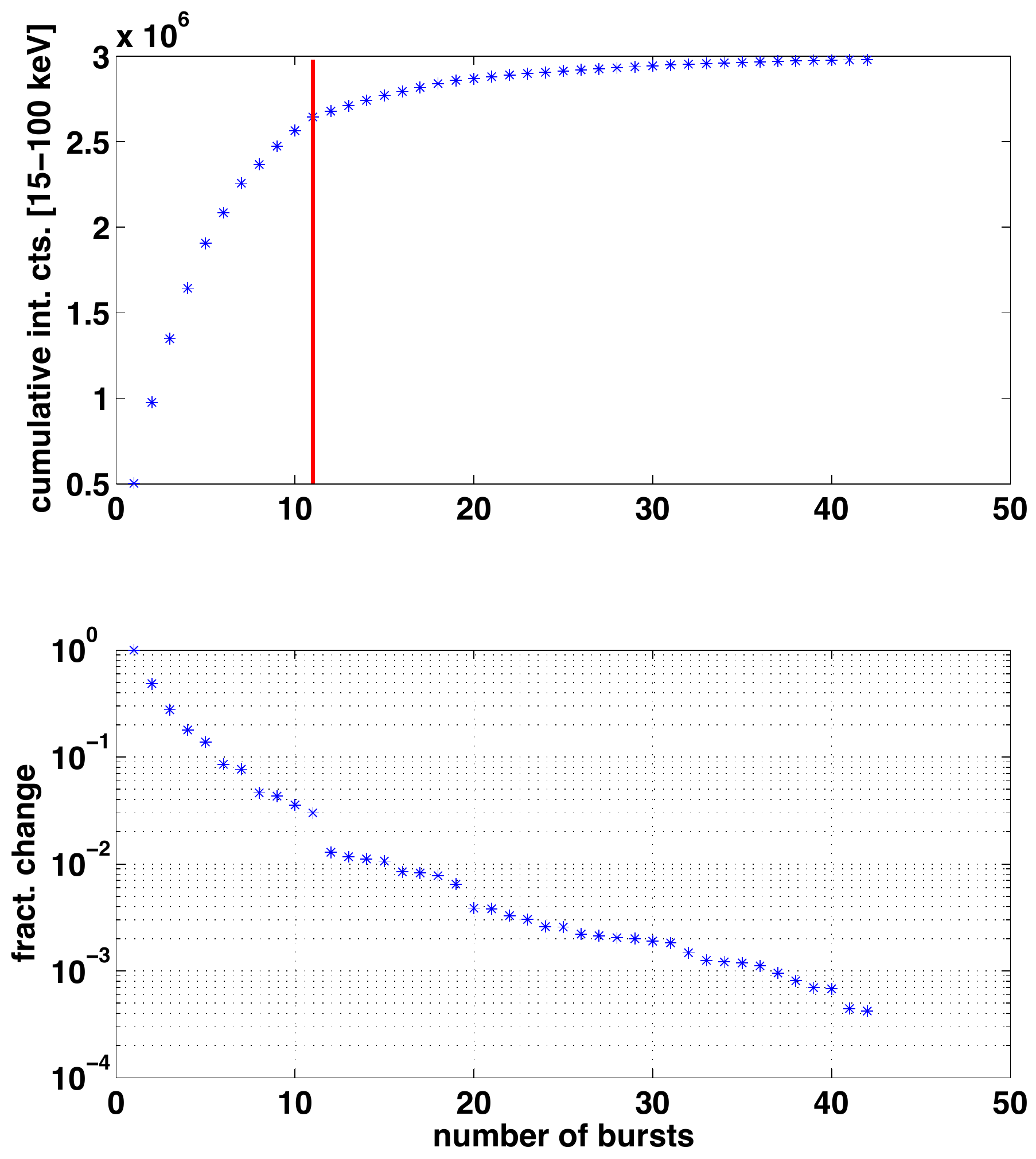}
\caption[] { Top plot: cumulative integrated counts.  The vertical line marks the contribution from the 11th most 
energetic peak.  89\% of the total integrated counts in the 42 storm bursts measured is contained in the most energetic 
11 bursts.  The twelfth most energetic burst contributes an additional 1\% to the total estimated from the 42 bursts.  
Bottom plot: fractional change in the cumulative integrated counts.  The 16th burst contributes less than 1\% to the 
running total.  The 37th burst contributes less than 0.1\% to the running total.  } \label{fig:cumulativeFluence}
\end{center}
\end{figure}

We chose to explore two GW emission stacking models in this mock search:  an $N=11$ flat model which sets GW energy 
emission constant; and a fluence-weighted model comprised of the 18 brightest bursts in the light curve which sets 
$\gamma$ to be constant.  The $N=11$ cutoff in the flat model (making it a step function) is motivated by burst integrated counts (Figure\,
\ref{fig:fluenceHist}).   The $N=18$ cutoff in the fluence-weighted model was motivated by Figure\,
\ref{fig:cumulativeFluence}.    Including the most energetic 18 bursts accounts for 95\% of the counts in the 42 bursts 
considered for the analysis.  After the 15 most energetic bursts, each additional burst (when ordered by energy)  
contributes less than 1\% to the total.  In Monte Carlo simulations we observed only a 3\% difference (averaged over 
amplitude sensitivity estimates set via the 12 injection waveforms) between fluence-weighted models with cutoffs at 
$N=17$ and $N=35$.

We chose a fluence-weighted GW emission model to explore the constant-$\gamma$ hypothesis.  We chose not to pursue 
GW emission models weighted to e.g. burst peak heights or burst rising edge slopes.   The cross-correlation between estimated 
burst peak counts and burst integrated counts is 0.75 in a normalization where auto-correlations are unity.  In the 
fluence-weighted emission model, assuming that GW emission energy is proportional to fluence we weight compound 
simulations with the square root of integrated counts, and then normalize so that the weight of the most energetic burst 
is unity.  We weight significance tilings on the other hand according to burst integrated counts before stacking them.

\subsection{Results}

In this section we present the results of a mock search with the P-Stack pipeline for GW associated with the 2006 
March 29 SGR 1900+14 storm, using simulated data.  At the time of the storm, all three LIGO detectors were taking 
science-quality data.  Simulated data modeled on data from the two LIGO 4\,km detectors were created from 
gaussian noise by matching power spectra with the LIGO data in the frequency domain.  Therefore, the sensitivity estimates 
should be similar to upper limit estimates from real data.  

\begin{figure}[!t]
\begin{center}
\includegraphics[angle=0,width=80mm, clip=false]{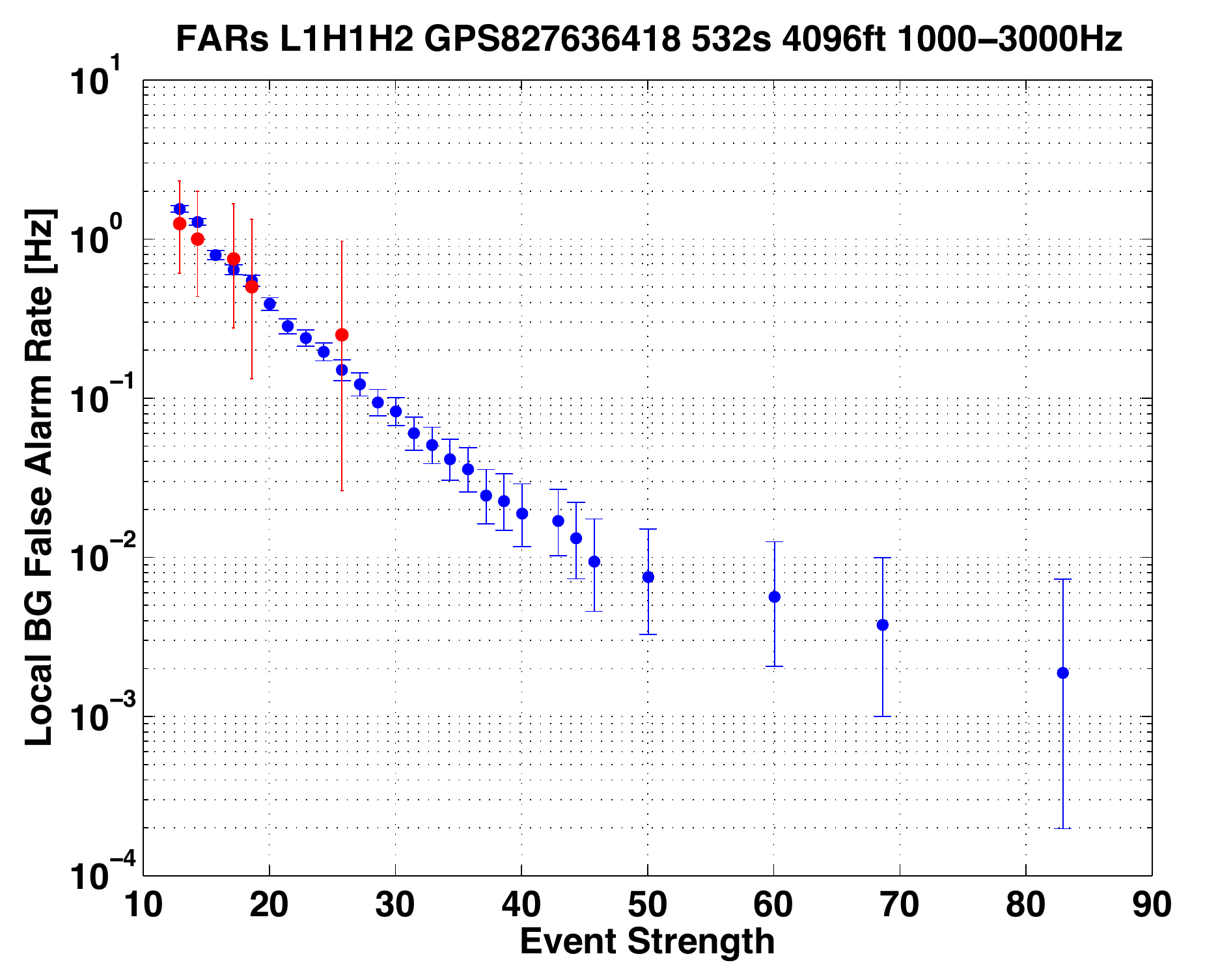}
\caption[Rate versus analysis event loudness] { Example cumulative histograms showing false alarm rates versus 
analysis event loudness for the background (blue) and the stacked on-source region (red).  There are three such plots 
for each emission model, one per search band.  Since the stacked livetime is 4\,s here, the loudest on-source event occurs 
once per 4\,s, and is plotted at a y-value of 0.25\,Hz.    We can estimate the FAR of this loudest on-source analysis 
from the background.  Error bars are Poissonian at 90\% confidence.}
\label{fig:exampleStackFAR}
\end{center}
\end{figure}

In Figure\,\ref{fig:exampleStackFAR} we show example cumulative histograms showing false alarm rates versus 
analysis event loudness for the background and the stacked on-source region.  There are three such plots for each 
emission model, one per search band.  Since the stacked livetime is 4\,s here, the loudest on-source event occurs once per 
4\,s, and is plotted at a y-value of 0.25\,Hz.    We can estimate the FAR of this loudest on-source analysis from the 
background.  

If only one emission model yields a significant event, or if the case is marginal,  we can gain additional information by making a joint statement of probability.   We can examine the background probability density function in a 2-dimensional space comprised of duples $(A,B)$ of loudest events from corresponding background segments analyzed under emission models $A$ and $B$, and determine constant probability contours $c=\alpha A + B$, where $\alpha$ is a normalizing constant found with a linear fit of the background scatter plot constrained to pass through the origin.  These contours then assign a joint probability to the loudest event duple $(a_L,b_L)$. 

\begin{figure}[!t]
\begin{center}
\includegraphics[angle=0,width=85mm, clip=false]{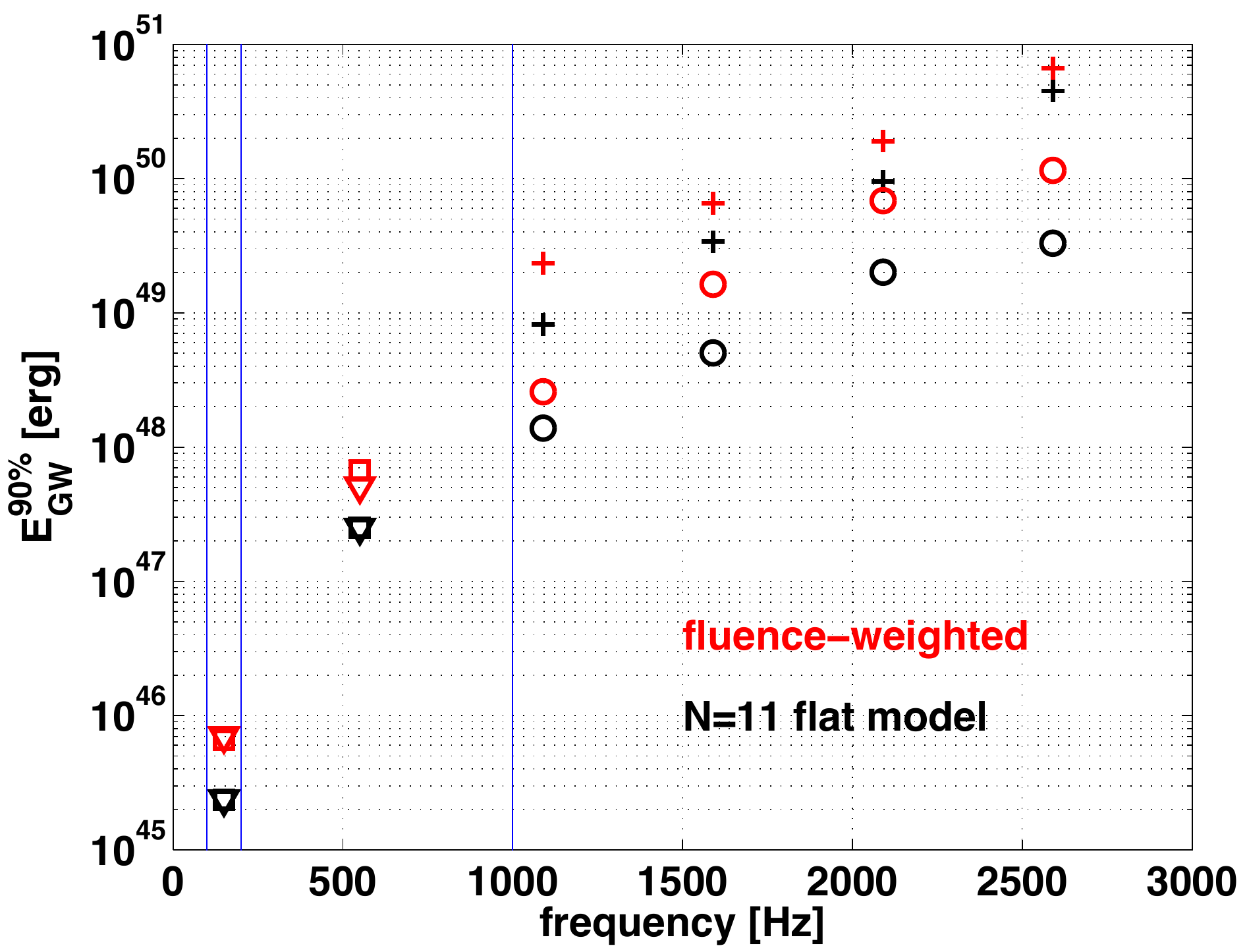}
\caption[Stack-a-flare simulated data energy sensitivity estimates] { Stack-a-flare simulated data energy sensitivity 
estimates at 10\,kpc, for the 29 March 2006 storm from SGR 1900+14, for the flat and  fluence-weighted models.  
Uncertainty estimates have been folded in, as tabulated in Table\,\ref{table:stacksim}.  Vertical lines indicate boundaries of the three distinct search frequency bands. Crosses and circles indicate linearly and circularly polarized RDs, respectively.  Triangles and 
squares represent 11\,ms and 100\,ms band- and time-limited WNBs, respectively.  Symbols are placed at the waveform central frequency.  These sensitivity estimates reflect the noise curves of the detectors. }
\label{fig:stacksimegw}
\end{center}
\end{figure}

Table\,\ref{table:stacksim} shows sensitivity estimates on GW amplitude and GW isotropic emission energy assuming 
a distance of 10\,kpc to SGR 1900+14, at 90\% detection efficiency, for the $N=11$ flat model and the fluence-
weighted model.  Sensitivity estimates from simulated data (in which there could be no GW signal) are produced 
using the identical procedure for producing upper limit estimates in a real search.  Sensitivity estimates with $N=1$ 
are also shown for the sake of comparison; when $N=1$ the Stack-a-flare pipeline reduces to the individual burst 
search pipeline (Flare pipeline).   Superscripts in Table\,\ref{table:stacksim} give a systematic error and uncertainties 
at 90\% confidence. The first and second superscripts account for systematic error and statistical uncertainty in 
amplitude and phase of the detector calibrations, estimated via Monte Carlo simulations, respectively. The third is a 
statistical uncertainty arising from using a finite number of injected simulations, estimated with the bootstrap method 
using 200 ensembles\,\cite{efron79}.  The systematic error and the quadrature sum of the statistical uncertainties are 
added to the final sensitivity estimates.   Methods used to estimate these uncertainties are further described in\,
\cite{kalmus08}.   As mentioned in Section\,\ref{section:light curve},  one-sigma burst timing uncertainties as measured 
by fits of burst rising edges are built into the Monte Carlo.   We  present the energy sensitivity estimates graphically in 
Figure\,\ref{fig:stacksimegw}.

We also present sensitivity estimates on $\gamma \equiv \egwn / \eem$,  a measure of the extent to which an energy 
upper limit probes the GW emission efficiency.   Estimates of $\gamma$, being ratios of energies, do not depend on 
the distance to SGR 1900+14.

\section{Discussion and conclusion}

We have presented the Stack-a-flare method for searching for GW associated with multiple SGR bursts that extends the individual 
SGR burst search presented in\,\cite{s5y1sgr, kalmus08}.   We have characterized both the T-Stack and the P-Stack 
versions of the Stack-a-flare pipeline, demonstrating sensitivity dependence on stacking number $N$ and uncertainty 
in relative timing between bursts.  The P-Stack pipeline is robust to timing errors, and we have used it to estimate GW 
amplitude, isotropic emission energy, and $\gamma$ search sensitivities for a mock SGR 1900+14 storm multiple 
SGR search, using simulated data modeled after LIGO data from the science run which was ongoing at the time of the SGR storm.    We considered two 
GW emission models to inform the stacking: flat with $N=11$; and fluence-weighted with $N=18$. 

Two other searches for GWs associated with SGR events, besides\,\cite{s5y1sgr}, have been published previously; 
neither claimed detection.  The AURIGA collaboration searched for GW bursts associated with the SGR 1806$-20$ 
giant flare in the band 850--950~Hz with damping time 100~ms, setting upper limits on the GW energy of $
\sim10^{49}$~erg\,\cite{auriga05}. The LIGO collaboration also published on the same giant flare, targeting times and 
frequencies of the quasi-periodic oscillations in the flare's x-ray tail as well as other frequencies in the detector's band, 
and setting upper limits on GW energy as low as $8\times10^{46}$~erg for quasi-periodic signals lasting tens of 
seconds\,\cite{matone07}.

Based on the pipeline characterization, we expect to gain a factor of $\sqrt{11}=3.3$ in energy sensitivity in the 
$N=11$ flat model over the $N=1$ case.  We observe a gain of 3.8 averaged over results obtained with the 12 
injection waveforms.   Comparing these estimated sensitivities to the individual burst search results for the SGR 
1900+14 storm published in\,\cite{s5y1sgr}, which analyzed the entire storm in a single $\pm20$\,s on-source region, 
we observe a gain in energy sensitivity of an order of magnitude.  The best values of $\gamma$ in\,
\cite{s5y1sgr}, for the 2004 December SGR 1806--20 giant flare, are in the range $\sci{5}{1}$--$\sci{6}{6}$ depending 
on injection waveform.  These $\gamma$ upper limits are about a factor of $10^3$ lower than the estimated $\gamma
$ sensitivities presented here.  However, the Advanced LIGO detectors promise an improvement in $\hrss$ by more 
than a factor of 10 over S5, corresponding to an improvement in energy sensitivity by more than a factor of 100.  A 
stacking search similar to this mock search in Advanced LIGO data could thus conceivably beat the SGR 1806--20 
giant flare $\gamma$ upper limits.  Furthermore, on 2008 August 22, a new SGR was discovered\,\cite{gcn8112, 
gcn8113, gcn8115} which may be located at a distance of only 1.5\,kpc in the direction of the galactic anti-center\,
\cite{gcn8149, leahy95}.  A stacking analysis on bursts from this SGR could therefore gain almost 2 additional orders 
of magnitude in energy and $\gamma$ upper limits.

This method would be suitable for a multiple burst search for GW associated with the SGR 1900+14 storm using 
real LIGO data.  A real search would be similar to the mock search in simulated data presented here, although detector collaborations may 
choose to explore different or additional stacking emission models.    This method would also be suitable for multiple SGR burst searches 
on isolated bursts spanning months or years.   We expect this method to continue to be useful when advanced detectors with greater sensitivity come on line.

\begin{acknowledgments}

LIGO was constructed by the California Institute of Technology
and Massachusetts Institute of Technology with funding
from the National Science Foundation and operates under
cooperative agreements PHY-0107417 and PHY-0757058.  The authors are also grateful for the support of the National
Science Foundation under grants PHY-0457528/0757982, PHY-0555628,  the California Institute of Technology, Columbia University in the City of New York, and the Pennsylvania State University.  We are indebted to many of our colleagues for fruitful discussions, in particular Richard O'Shaughnessy for his valuable suggestions.  This paper
has been assigned LIGO Document Number LIGO-\ligodoc.

\end{acknowledgments}

\begin{sidewaystable*}
\caption[Sensitivity estimates for a simulated SGR 1900+14 storm stacking search ]{Sensitivity estimates for a 
simulated SGR 1900+14 storm search in simulated data, for two GW emission stacking models ($N=11$ flat model and 
fluence-weighted model).  The $N=1$ case is shown for comparison.  Results are shown for various ringdown and 
band- and time-limited white noise burst target signal classes.   One-sigma burst timing uncertainties as measured by 
fits of burst rising edges are included in the Monte Carlo. Superscripts give a systematic error and uncertainties at 90\% confidence. Similar estimates were made for the $\egwn$ sensitivity estimates, but are not shown in the table.   The 
first and second superscripts account for systematic error and statistical uncertainty in amplitude and phase of the 
detector calibrations used to create the data from which the simulated data ws modeled, estimated via Monte Carlo 
simulations, respectively. The third is a statistical uncertainty arising from using a finite number of injected simulations, 
estimated with the bootstrap method using 200 ensembles.  The systematic error and the quadrature sum of the 
statistical uncertainties are added to the final sensitivity estimates.  Sensitivity estimates on $\gamma \equiv \egwn / 
\eem$ are also given, using a conservative estimate of $\sci{1.0}{-4}$\,erg cm$^{-2}$ for the total fluence of the storm 
to estimate fluences for individual peaks. }
\begin{scriptsize}
\begin{tabular}{@{\extracolsep{\fill}}l||rlr|rr||rlr|rr||rlr|rr}
 \hline \hline
 & \multicolumn{5}{c}{N=1} & \multicolumn{5}{c}{N=11 Flat} & \multicolumn{5}{c}{Fluence-weighted} \\ 
 Simulation type & \multicolumn{3}{c}{$ \hrssn [ 10^{-22}~ \rthz $] }  & $\egwn$ [erg] & $\gamma$  & \multicolumn{3}{c}
{$ \hrssn [ 10^{-22} ~ \rthz $] }  & $\egwn$ [erg] & $\gamma$ & \multicolumn{3}{c}{$ \hrssn [ 10^{-22} ~ \rthz $] }  & $
\egwn$ [erg] & $\gamma$ \\ 
 \hline 
 WNB 11ms 100-200 Hz   & 3.9 & $^{ +0.0 ~ +0.40 ~ +0.43}$ &  $= 4.5$ & $\sci{1.9}{46}$ & $\sci{2}{05}$ & 1.4 & 
$^{ +0.0 ~ +0.14 ~ +0.11}$ &  $= 1.6$ & $\sci{2.3}{45}$ & $\sci{5}{04}$ & 2.3 & $^{ +0.0 ~ +0.24 ~ +0.30}$ &  $= 2.7$ & 
$\sci{6.9}{45}$ & $\sci{8}{04}$ \\
 WNB 100ms 100-200 Hz   & 2.9 & $^{ +0.0 ~ +0.30 ~ +0.33}$ &  $= 3.3$ & $\sci{1.0}{46}$ & $\sci{1}{05}$ & 1.4 & 
$^{ +0.0 ~ +0.15 ~ +0.13}$ &  $= 1.6$ & $\sci{2.3}{45}$ & $\sci{5}{04}$ & 2.2 & $^{ +0.0 ~ +0.23 ~ +0.36}$ &  $= 2.6$ & 
$\sci{6.5}{45}$ & $\sci{7}{04}$ \\
 WNB 11ms 100-1000 Hz   & 6.6 & $^{ +0.0 ~ +0.68 ~ +0.79}$ &  $= 7.6$ & $\sci{7.6}{47}$ & $\sci{8}{06}$ & 4.0 & 
$^{ +0.0 ~ +0.42 ~ +0.42}$ &  $= 4.6$ & $\sci{2.4}{47}$ & $\sci{6}{06}$ & 5.8 & $^{ +0.0 ~ +0.60 ~ +0.54}$ &  $= 6.6$ & 
$\sci{5.0}{47}$ & $\sci{5}{06}$ \\
 WNB 100ms 100-1000 Hz   & 6.3 & $^{ +0.063 ~ +0.66 ~ +0.85}$ &  $= 7.5$ & $\sci{6.2}{47}$ & $\sci{7}{06}$ & 4.1 & 
$^{ +0.041 ~ +0.43 ~ +0.29}$ &  $= 4.7$ & $\sci{2.5}{47}$ & $\sci{6}{06}$ & 6.5 & $^{ +0.065 ~ +0.67 ~ +0.98}$ &  $= 
7.8$ & $\sci{6.6}{47}$ & $\sci{7}{06}$ \\
 RDC 200ms 1090 Hz   & 9.9 & $^{ +0.20 ~ +1.0 ~ +1.3}$ &  $= 12$ & $\sci{5.9}{48}$ & $\sci{6}{07}$ & 4.7 & 
$^{ +0.095 ~ +0.49 ~ +0.55}$ &  $= 5.6$ & $\sci{1.4}{48}$ & $\sci{3}{07}$ & 6.2 & $^{ +0.12 ~ +0.64 ~ +1.2}$ &  $= 7.7$ 
& $\sci{2.6}{48}$ & $\sci{3}{07}$ \\
 RDC 200ms 1590 Hz   & 13 & $^{ +0.53 ~ +1.4 ~ +3.1}$ &  $= 17$ & $\sci{2.9}{49}$ & $\sci{3}{08}$ & 6.2 & $^{ +0.25 
~ +0.64 ~ +0.68}$ &  $= 7.4$ & $\sci{5.0}{48}$ & $\sci{1}{08}$ & 11 & $^{ +0.44 ~ +1.1 ~ +1.4}$ &  $= 13$ & $\sci{1.6}
{49}$ & $\sci{2}{08}$ \\
 RDC 200ms 2090 Hz   & 23 & $^{ +1.8 ~ +3.0 ~ +3.6}$ &  $= 30$ & $\sci{1.3}{50}$ & $\sci{1}{09}$ & 9.0 & $^{ +0.72 ~ 
+1.2 ~ +0.61}$ &  $= 11$ & $\sci{1.9}{49}$ & $\sci{4}{08}$ & 16 & $^{ +1.3 ~ +2.1 ~ +2.1}$ &  $= 20$ & $\sci{6.7}{49}$ 
& $\sci{7}{08}$ \\
 RDC 200ms 2590 Hz   & 20 & $^{ +2.6 ~ +2.6 ~ +2.5}$ &  $= 26$ & $\sci{1.6}{50}$ & $\sci{2}{09}$ & 9.3 & $^{ +1.2 ~ 
+1.2 ~ +0.39}$ &  $= 12$ & $\sci{3.4}{49}$ & $\sci{8}{08}$ & 17 & $^{ +2.3 ~ +2.2 ~ +1.8}$ &  $= 23$ & $\sci{1.2}{50}$ 
& $\sci{1}{09}$ \\
 RDL 200ms 1090 Hz   & 16 & $^{ +0.48 ~ +1.6 ~ +3.6}$ &  $= 20$ & $\sci{1.7}{49}$ & $\sci{2}{08}$ & 10 & $^{ +0.31 ~ 
+1.1 ~ +2.8}$ &  $= 14$ & $\sci{8.4}{48}$ & $\sci{2}{08}$ & 16 & $^{ +0.48 ~ +1.7 ~ +6.5}$ &  $= 23$ & $\sci{2.4}{49}$ 
& $\sci{3}{08}$ \\
 RDL 200ms 1590 Hz   & 32 & $^{ +1.6 ~ +3.3 ~ +8.7}$ &  $= 43$ & $\sci{1.7}{50}$ & $\sci{2}{09}$ & 15 & $^{ +0.74 ~ 
+1.5 ~ +3.5}$ &  $= 19$ & $\sci{3.4}{49}$ & $\sci{8}{08}$ & 21 & $^{ +1.0 ~ +2.2 ~ +3.7}$ &  $= 26$ & $\sci{6.6}{49}$ & 
$\sci{7}{08}$ \\
 RDL 200ms 2090 Hz   & 35 & $^{ +3.8 ~ +4.5 ~ +17}$ &  $= 56$ & $\sci{5.5}{50}$ & $\sci{6}{09}$ & 18 & $^{ +2.0 ~ 
+2.3 ~ +3.3}$ &  $= 24$ & $\sci{9.6}{49}$ & $\sci{2}{09}$ & 23 & $^{ +2.5 ~ +2.9 ~ +11}$ &  $= 37$ & $\sci{1.9}{50}$ & 
$\sci{2}{09}$ \\
 RDL 200ms 2590 Hz   & 51 & $^{ +6.6 ~ +6.5 ~ +17}$ &  $= 76$ & $\sci{1.4}{51}$ & $\sci{2}{10}$ & 29 & $^{ +3.7 ~ 
+3.7 ~ +7.9}$ &  $= 41$ & $\sci{4.7}{50}$ & $\sci{1}{10}$ & 38 & $^{ +4.9 ~ +4.8 ~ +7.0}$ &  $= 51$ & $\sci{6.8}{50}$ & 
$\sci{7}{09}$ \\
 \hline \label{table:stacksim} 
 \end{tabular} 
 \end{scriptsize} 
 \end{sidewaystable*}

\end{document}